\documentclass{aa501}

\usepackage[dvips]{graphicx}
\usepackage[normalem]{ulem}

\newcommand\vect[1]{{\mbox{\boldmath $#1$}}}
\newcommand{\crit}{{{\rm cr}}}
\newcommand{\lprop}{\;
  \raise0.3ex\hbox{$\propto$\kern-0.75em\raise-1.1ex\hbox{$\sim$
  }}\;\hskip-2pt }
\newcommand{\HI}{{\ion{H}{i}}}
\newcommand{\HII}{{\ion{H}{ii}}}
\newcommand{\B}{{B}}
\newcommand{\nel}{{n_{\rm e}}}

\newcommand{\kms}{\,{\rm km\,s^{-1}}}
\newcommand{\cm}{\,{\rm cm}}
\newcommand{\cmcube}{\,{\rm cm^{-3}}}
\newcommand{\G}{\,{\rm G}}
\newcommand{\gcmcube}{\,{\rm g}\,{\rm cm^{-3}}}

\newcommand{\mkG}{\,\mu{\rm G}}

\newcommand{\kpc}{\,{\rm kpc}}

\newcommand{\yr}{\,{\rm yr}}
\newcommand{\s}{\,{\rm s}}

\newcommand{\radm}{\,{\rm rad\,m^{-2}}}

\begin{document}

\title{Magnetic fields in barred galaxies}
\subtitle{II. Dynamo models}
\author{D.~Moss$^1$ \and A.~Shukurov$^2$ \and D.~Sokoloff$\,^3$ \and
R.~Beck$^4$
\and A.~Fletcher$^2$
}

\offprints{D.~Moss}

\institute{
Department of Mathematics, University of Manchester, Manchester M13 9PL, UK
\and
Department of Mathematics, University of Newcastle,  Newcastle on Tyne NE1 7RU, UK
\and
Department of Physics, Moscow State University, Moscow 19899, Russia
\and
Max Planck Institut f\"ur Radioastronomie, Auf dem H\"ugel 69, 53121 Bonn, Germany
}

\date{Received ...; accepted ...}

\authorrunning{Moss et al.}

\titlerunning{Magnetic fields in barred galaxies. II. Dynamo models}
\abstract{
  We study the generation and maintenance of large-scale magnetic
  fields in barred galaxies. We take a velocity field (with strong
  noncircular components) from a published gas dynamical simulation
  of Athanassoula (1992), and use this as input to a galactic dynamo
  calculation.  Our work is largely motivated by recent high quality
  VLA radio observations of the barred galaxy NGC~1097, and we
  compare our results in detail with the regular magnetic fields
  deduced from these observations. We are able to reproduce most of
  the conspicuous large-scale features of the observed regular field,
  including
  the field structure in the central regions, by using a simple
  mean-field dynamo model in which the intensity of interstellar
  turbulence (more precisely, the turbulent diffusivity) is enhanced
  by a factor of 2--6 in the dust lanes and near the circumnuclear
  ring.  We argue that magnetic fields can be dynamically important,
  and therefore should be included in models of gas flow in barred
  galaxies.
 }

\maketitle

\section{Introduction}                  \label{intro}
\begin{figure}
\centerline{\includegraphics[width=0.99\hsize]{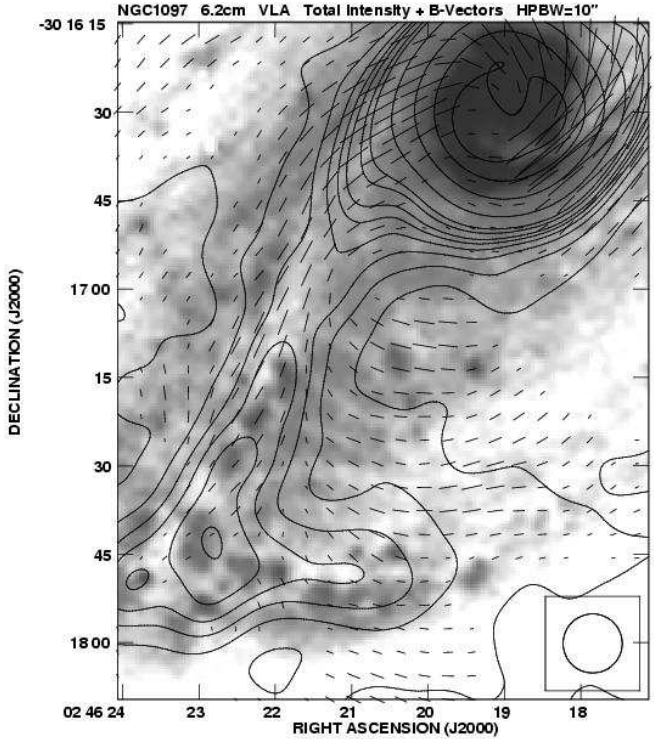}}
\caption{\label{N1097obs}The radio continuum emission
in the southern part of NGC~1097, with the galactic centre
  at top right, observed at
  $\lambda6.2\cm$ with the VLA (see Paper III for details).  Contours
  give the total intensity (representing the total magnetic field),
  dashes show the observed polarization plane rotated by $90^\circ$
  (representing the regular magnetic field uncorrected for Faraday rotation).
The Faraday rotation is about
  $\pm20^\circ$ at this wavelength
  (although it can be larger in localized regions).  The
  length of the dashes is proportional to the polarized intensity.
  The half-power width of the synthesized beam (shown in the lower
  right corner) is 10\arcsec.  The background optical B-band image
  (grey scale) was obtained at the Cerro Tololo Interamerican
  Observatory by H.~Arp.  The bar major axis is at an angle
  $-30^\circ$ and the disc major axis (the line of nodes) is at
  $-45^\circ$, both with respect to the south-north direction.
  The south-western side of the galaxy is
  closer to us. }
\end{figure}

Barred galaxies possess remarkably rich and unusual magnetic
structures. The first systematic observations of polarized radio
emission from a sample of barred galaxies (Beck et al.\ 2001a,
hereafter Paper I; see also Beck et al.\ 1999) have revealed
widespread polarized emission in the bar region, indicative of
magnetic fields ordered at scales of order 1\,kpc. The best examples
to date are the nearby prototypical barred galaxies NGC~1097 (Beck et
al.\ 1999, see Fig.~\ref{N1097obs}) and NGC~1365 (Paper I; see also
Fig.~7 in Lindblad 1999), and these galaxies show broadly similar
magnetic structures.

The field in the bar region, obtained from the $B$-vector of the
polarized radio emission at centimeter wavelengths where Faraday
rotation is small, appears approximately to be aligned with the gas
streamlines resulting from gas dynamical simulations (Athanassoula
1992; henceforth A92); notably, it is the velocity field in the
reference frame corotating with the bar that exhibits the alignment.
Such an alignment is perhaps not surprising in view of the strong gas
velocity shear (Beck et al.\ 1999).
In NGC~1365 (Lindblad et al.\
1996) and NGC~1530 (Reynaud \& Downes 1998) the gas
has strong
velocity gradients near the dust lanes, consistent with the predictions of
gas dynamical models.
For NGC~1097, velocity data with sufficient resolution are not
available.
 The total
nonthermal radio emission,
a tracer of the total (regular+turbulent) magnetic field,
exhibits ridges near the dust lanes (see
also Ondrechen \& van der Hulst 1983) that occur because of the strong
gas compression downstream of the bar's major axis (A92).  Thus, the
hydrodynamic shock appears to have a counterpart in the total magnetic
field (but not in the regular field -- see below).

There are, however, a number of puzzling features in the magnetic
field which are especially evident in the regular magnetic field (as
opposed to the turbulent field whose scale is of order 100\,pc).
Firstly, the contrast in the total magnetic field strength near the
dust lanes is significantly lower (about 2 in NGC~1097, see Beck et
al.\ 1999) than that in gas density (about 20 or more), as inferred from models
(A92, Englmaier \& Gerhard 1997)
and from CO observations (e.g.\ for     NGC~1530, Reynaud \& Downes 1998).
This suggests that the total magnetic field strength $\B$ does not
scale with the local gas density $\rho$, as would follow from a
one-dimensional shock compression in the dust lanes, $\B\propto\rho$,
and even $\B\propto\rho^{1/2}$ is too strong a dependence.
Beck et al.\ (1999) argue that the low magnetic field
compression could be due to enhanced turbulent diffusion downstream
from the shock, but this idea needs quantitative verification.

Moreover, the polarization intensity (a measure of the strength of the
regular magnetic field) in NGC~1097 has a broad local maximum upstream
of the dust lanes, where gas density has deep minimum.
The regular magnetic field shows hardly any enhancement in the dust lanes.
It is also
surprising that the regular magnetic field does not show any
discontinuity associated with the shock. The sharp turn in magnetic
lines leading to a depolarized strip detected by Beck et al.\ (1999)
has been resolved in later observations (Beck et al. 2001b, Paper III)
into a smooth turning of the magnetic lines in the region 1\,kpc
upstream of the dust lanes (Fig.~\ref{N1097obs}).  The turn is even
smoother in NGC~1365 and occurs about 3\,kpc upstream of the
compression region (Paper III). Thus, there is no clear signature of
{\it any\/} shock-related compression
or sharp deflection in the regular magnetic field.
It appears that shock effects in the magnetic field may have been
mostly eradicated by the
enhanced turbulent diffusion in the sheared shock region. The
verification and quantification of this effect is one of the main
goals of this paper.

There are thin, feather-like, dust filaments bent in a similar way
upstream of the main dust lane, seen in the optical photographs of
NGC~1097 and NGC~1365. The observed magnetic field is well aligned
with these filaments, especially at a distance of 1--3\,kpc upstream of the shock
front in NGC~1097 (see Fig.~\ref{N1097obs}). Similar
filaments have been observed in NGC~5383 by Sheth et al.\ (2000) who
observe that they are associated with \HII\ regions and propose that
the high density and low shear in the filaments make them hosts to
star forming regions. If the dust filaments trace the
streamlines, the velocity vectors must deviate from the streamlines
predicted for this region by gas dynamical simulations (see e.g.\
Fig.~12 in Roberts et al.\ 1979, Fig.~2b in A92, and
Figs~16 and 26 in Lindblad et al.\ 1996 where the streamlines have sharp
cusp rather than turn smoothly upstream of the shock front). If the
magnetic field is sufficiently strong in that region it may control
the shape of the dust filaments; however, the occurrence of a
dynamically dominant magnetic field has to be confirmed
observationally and then explained.

These facts imply that the physics involved is more complicated than
that of a simple shock with passive magnetic field.
The equipartition strength (with respect to the cosmic rays) of the
observed regular magnetic field is
about $7\mkG$ upstream of the dust lane in the southern part of
NGC~1097 (Beck et al.\ 1999). Taking a gas number density of
$n=0.1$--$0.2\cmcube$ (from A92), we obtain an Alfv\'en speed of
30--$50\kms$. This is much larger than the speed of sound (and,
presumably, the turbulent velocity if it has the standard value of
$10\kms$) and is comparable to the shearing velocity in most of the
bar region. Hence the magnetic
energy density seems to be comparable to the kinetic energy density in
the regular shearing motion and much exceeds the thermal energy
density. It follows that a fully consistent model of gas flow in barred galaxies
should include magnetic fields,
a factor systematically neglected in all gas
dynamical models of barred galaxies. In this paper we discuss models
of regular magnetic field evolution in the velocity field resulting
from Athanassoula's (1992) gas dynamical model in order to assess the
extent of the problem, i.e.\ to establish which features of the
observed magnetic field can be readily explained by existing knowledge
of the gas flow in barred galaxies and which appear to demand more
advanced MHD modelling.

Models of magnetic field evolution in barred galaxies, without the
inclusion of dynamo action, have been considered by
Otmianowska-Mazur et al.\ (1997), who used a velocity field resulting
from $N$-body simulations (with 38,000 stars and 19,000 molecular
clouds). As usual with $N$-body simulations with a modest number of
particles, their density distribution has only
a moderate enhancement near the bar major axis where dust lanes
occur in real galaxies. The simulated magnetic field concentrates in
the spiral arms outside the bar region and is weak in the bar. Since
the magnetic field is not supported by any dynamo action, it decays in
less than $5\times10^8\yr$.

Dynamo models for barred galaxies have been considered by Moss et al.\
(1998, 1999) who used the gas velocity
field obtained from $N$-body simulations with 200,000 stars and 40,000
gas particles. Their density field shows appropriately stronger
density contrasts and the inclusion of dynamo action results in
persistent magnetic structures. However, the distribution of particles
in such models is sparse and uneven, so it is difficult to define
reliably a velocity field in low-density regions.  This leads to
serious complications in magnetic field modelling, and so we prefer to
use a velocity field obtained from gas dynamical simulations.

Our model focuses on the regular magnetic field, and the effects of
interstellar turbulence on the magnetic field are parameterized in
terms of turbulent transport coefficients, including a standard
$\alpha$-effect (Moffatt 1978, Parker 1979,  Krause \& R\"adler
1980). The formulation would be
similar for alternative mechanisms, such as a magnetic buoyancy
driven dynamo, as suggested by Moss et al.\ (1999b).  The key feature
of our model is the inclusion of a realistic representation at high
spatial resolution of the velocity field in barred galaxies.

We also report briefly in Sect.~\ref{extra} on our discovery of a
rather remarkable dynamo effect in galactic discs with positive dynamo
number,
where the dynamo is driven by noncircular motions rather than by
differential rotation. This may be of interest in the wider
context of dynamo theory.

\section{The alignment of magnetic and velocity fields}
\begin{figure}
\centerline{\includegraphics[width=0.99\hsize]{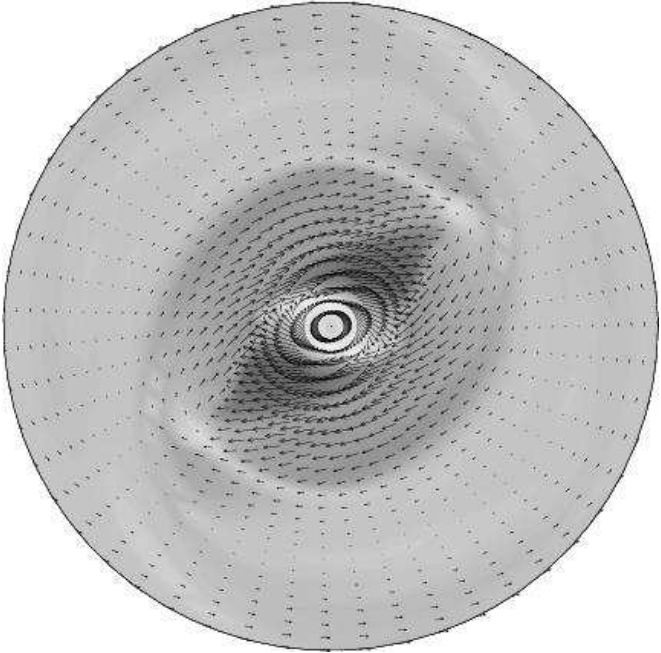}}
\caption{\label{vel0}The velocity and gas density (grey scale, lighter shades correspond to
higher values) fields  in the frame
corotating with the bar, from
Model 001 of A92. The sense of rotation is
clockwise.  The bar major axis is at an angle of $45^\circ$ to the
vertical, the frame radius is $8\kpc$ and the corotation
radius is $r_{\rm c}\approx6\kpc$.
}
\end{figure}

As noted by Beck et al.\ (1999), the regular magnetic field in the bar
region of NGC~1097 seems to be aligned with the velocity field of A92,
especially in regions with stronger velocity shear.  This can be seen
from a comparison of Fig.~\ref{N1097obs}, where we present the
polarization map of NGC~1097, with the magnetic field orientation
indicated with dashes, and Fig.~\ref{vel0} where a velocity field from
A92 is shown.  Such an alignment is not typical of normal spiral
galaxies where magnetic field lines are inclined to the streamlines by
$10^\circ$--$30^\circ$, presumably due to the dynamo action (e.g.,
Beck et al.\ 1996, Beck 2000, Shukurov 2000).
In the presence of
a strongly sheared velocity, the local structure of the magnetic field
will be controlled by the local velocity shear. In barred galaxies,
the shear of the noncircular velocity field is strong enough to make
the form of their magnetic fields markedly different from those found
in normal spiral galaxies.

Ignoring the effects of dynamo action, the large-scale field will be
frozen into the flow in regions
with $R_{\rm m}=uL/\eta\ga1$, where
$u\simeq 100\kms$ is the regular shearing velocity, $L\simeq 3\,\kpc$
is its scale in the
bar region, and $R_{\rm m}$ is the magnetic
Reynolds number based on the turbulent magnetic diffusivity $\eta$.
Thus, the field will be aligned with the flow if $\eta\la
10^{29}\cm^2\s^{-1}$.

However, the alignment of magnetic field and the flow can be affected
by dynamo action even at large values of $R_{\rm m}$.
Dynamo action is needed
to maintain the global magnetic field against the effects of winding
by differential rotation and tangling by turbulence, which would lead
eventually to enhanced Ohmic decay.
Therefore, we
also require that the dynamo is unable to misalign the field and the
streamlines: the local growth rate of the magnetic field $\gamma$ must
be smaller than the shear rate, i.e. $\gamma\la u/L$ with
$\gamma\simeq D^{1/2}\eta/h^2$, where $D=\alpha uh^3/(L\eta^2)$ is the
local dynamo number and $\alpha$ is the alpha-coefficient.  This
yields $\alpha L/uh\la1$, or
$Lv_{\rm t}/hu\la 1$, where $v_{\rm t}$
is the turbulent velocity and
$\alpha\leq v_{\rm t}$ (since $\alpha$ cannot exceed the
turbulent speed -- see, e.g., Sect.~V.4 in Ruzmaikin et al.\
1988). In normal galaxies where $u$ is dominated by the streaming
velocity produced by the spiral pattern and so $L/h\simeq2$--3 and
$v_{\rm t}/u\simeq 1$, this
{\bf inequality}
is not satisfied and we can not
expect strong alignment between the streamlines and magnetic lines.
Indeed, the magnetic pitch angle (i.e. the angle between the regular
magnetic field and the streamlines) is about 1/3 radian,
plausibly consistent with this estimate.
On the other hand, the shear rate $u/L$ is significantly larger in
barred galaxies and we can expect a much closer alignment in the
regions where $L\la3\kpc$ and $u\ga60\kms$, where the above
inequality is satisfied.  In other words, we expect that barred
galaxies contain large regions where dynamo action is overwhelmed by
the local velocity shear resulting in a tight alignment of the
magnetic field with the shearing velocity (more precisely, with the
principal axis of the rate of strain tensor).  On the other hand,
there may be regions of enhanced diffusivity and/or reduced shear
where the alignment is reduced (see Fig.~\ref{angle}).

The velocity field, but not the magnetic field, looks different in the
inertial and corotating frames.  We can expect a rather close general
alignment between $\vec B$ and $\vec u$ in the reference frame
corotating with the bar for the following reason.  Nonaxisymmetric
magnetic
field patterns must rotate rigidly to avoid winding up by
differential rotation (and it is the dynamo which can maintain such
fields). With a nonaxisymmetric perturbation from the bar (or spiral
arms), the
magnetic
modes that are corotating with the perturbation will be
preferentially excited (e.g.\
Mestel \& Subramanian 1991; Moss 1996, 1998; Rohde et al.\ 1999).
Thus, the regular magnetic field will corotate (or nearly corotate)
with the bar, and this is a physically distinguished reference frame.
(All magnetic field configurations discussed in this paper, except
those in Sect.~\ref{extra}, do exactly corotate with the bar.)  So,
an approximate alignment between $\vec{B}$ and $\vec{u}$ is expected
in the corotating, but not in the inertial, frame.

\section{The model}

\subsection{The gas velocity field}
\label{velfld}
The velocity field was taken from data supplied by
E.~Athanassoula, corresponding to Model 001 shown in Fig.~2 of A92.
This model is two-dimensional, with $\vec{u}=(u_x(x,y), u_y(x,y), 0)$
in cartesian coordinates $(x,y,z)$. The
gas velocity and density fields
are steady in the frame rotating with the bar (the corotating frame).
The stellar bar extends approximately between the ends of the dust
lanes (shown as a light shade in Fig.~\ref{vel0}).  We reproduce in
Fig.~\ref{vel0} the velocity field in this frame rotating in a
clockwise sense with angular velocity of about $34\kms\kpc^{-1}$,
placing the corotation radius at about $r=r_{\rm c}\approx6\kpc$.
The streaming velocities relative to the rotating frame are as large
as $100\kms$ near the shock fronts, which are slightly offset from
the major axis of the bar.  Regions where the sheared velocity is
comparable with the rotational velocity are widespread in the bar
region.
A shock near the dust lanes is just resolved in the data.  The region
in which the velocities are available is $16\kpc$ across.

The magnetic fields obtained using this
velocity field will be compared with observations of NGC~1097 where
the corotation radius is $r_{\rm c}\approx12\kpc$ (assuming a
distance to NGC~1097 of 17~Mpc); the length scales are therefore
renormalized correspondingly before any comparison is made.  However,
when discussing the computational results in Sects.~\ref{mainres} and
\ref{extra} we retain the original length scale
of A92 with corotation at $6\kpc$.
Since the model of A92 refers to a generic barred galaxy rather than to
the galaxy NGC~1097 specifically, our comparison with observations can only be
qualitative.

\subsection{The dynamo model}
\label{basic}
The basic dynamo model, as applied to galaxies with strong
noncircular motions, is described in Moss et al.\ (1999); it uses the
`no-$z$' approximation (Moss 1995), replacing derivatives
perpendicular to the galactic midplane ($z$-direction) with inverse
powers of $h$, the disc semi-thickness. This is in some ways
consistent with the two-dimensional flow model of A92.  The mean
field dynamo equations are solved for the magnetic field components
in the directions orthogonal to $z$. ($B_z$ is given in principle by
the condition $\nabla\cdot\vec{B}= \vec{0}$.) The resulting magnetic
field can be thought of as an approximation to the field in the
mid-plane, or to represent values averaged vertically over the disc.

We made one significant amendment to the model. Phillips (2001) has
showed that the dynamo growth rates in the `no-$z$' model can be made
to agree more closely with those of the asymptotic analysis (e.g.\
Ruzmaikin et al.\ 1988) by including a correction factor of $\pi^2/4$
into the terms representing diffusion in the vertical direction, and
so we also included these factors. Thus the local marginal dynamo numbers
for our models can now be expected to be fairly directly comparable to
standard values
of about 10.

The large velocity shear present in
barred galaxies resulted in
unsatisfactory numerical behaviour of the numerical algorithm
previously employed (e.g.\ Moss et al.\ 1998, 1999; Moss 1995), and so
a new version in cartesian coordinates was written, using
a second order Runge--Kutta method for time-stepping, and second-order
accurate space discretisation. This code solves the `no-$z$' dynamo
equations in the frame corotating with the bar. The earlier
calculations were carried out using the $\alpha\omega$ formulation,
but now we allow for the regeneration of both meridional and azimuthal
regular magnetic field by interstellar turbulence (the
$\alpha^2\omega$ dynamo). The dynamo equation has the form
\begin{equation}
\frac{\partial \vec{B}}{\partial
t}=\nabla\!\times\!(\vec{u}\times\vec{B}-
{\textstyle\frac{1}{2}}\nabla\eta\times\vec{B}+\alpha\vec{B})-
\nabla\!\times(\eta\nabla\!\times\!\vec{B}),
\label{mfd}
\end{equation}
where $\vec u$ is the regular velocity field described in
Sect.~\ref{velfld}, $\alpha$ parameterizes the dynamo action of the
interstellar turbulence, and $\eta$ is the turbulent magnetic
diffusivity (see, e.g., Ruzmaikin et al.\ 1988, Beck et al.\ 1996).
In Eq.~(\ref{mfd}), we ignore any vertical ($z$-wise) variation of
$\eta$, but allow a
variation in the disc plane. (In
principle we might, in the spirit of the no-$z$ approximation, have
approximated the effects of a halo diffusivity larger than that in the
disc by an estimate of the form $\partial\eta/\partial z \sim
\eta(x,y)/h$.) The equation is  solved in a cylindrical region with
radius $8\kpc$.  We allow for nonlinear dynamo effects leading to the
saturation of the magnetic field growth by adopting
\begin{equation}
\alpha=\alpha_0\frac{\tilde\alpha(\vec{r})}{1+\xi\vec{B}^2/4\pi\rho
(\vec{r})v_{\rm t}^2}\;,
\label{alphaq}
\end{equation}
where $\alpha_0$ (a typical amplitude of $\alpha$) is a constant,
$\tilde\alpha(\vec{r})$ is the normalized background value, and
$\rho$ and $v_{\rm t}$ are the gas density and turbulent velocity. We are
thus assuming that the large-scale magnetic field significantly
reduces the $\alpha$-effect when its energy density approaches that of
the turbulence;
$\xi$ is a constant,
introduced to suggest some of the uncertainty about the details of
this feedback. When estimating field strength
from our models we set $\xi=1$, but this fundamental uncertainty
should be kept in mind. (Indeed, it is quite possible that, rather
than being a constant, $\xi$ could depend on local conditions.) We
assume that $v_{\rm t}=10\kms$, and $\rho$ is taken from the same gas
dynamical model of A92 as the velocity field $\vec
u$. It is often assumed that $\alpha\propto\omega$ with $\omega$ the angular
velocity of rotation. We first consider models with $\alpha$ constant apart
from quenching effects, i.e.\ $\widetilde\alpha(\vec{r})=1$, and then put
$\widetilde\alpha(\vec{r})=\omega(r)/\omega(r_\omega)$, where we arbitrarily
choose $r_\omega=0.3R$, and $R=8\kpc$. (Note that
$\omega(r_\omega)/\Omega_0\approx 0.22$ where the unit angular velocity
$\Omega_0$ is introduced in Sect.~\ref{params}.) We choose this normalization
so as to have the dimensionless quantities of order unity in the outer parts of
the galaxy.

We can estimate the $z$-averaged vertical field in the disc, by using
the condition $\nabla \cdot \vec{B} = 0$, and estimating $\partial
B_z/\partial z$ by $B_z/h$. This  vertical magnetic
field (not shown in Fig.~\ref{model101} and other similar figures),
is small on average, being about 10 times weaker than the average
horizontal field. However, the vertical field is significant in
regions with pronounced structure in the horizontal field (e.g. the
shock front and the central region) where it can be comparable to the
horizontal field.

Note that the dynamo equations with the quadratic nonlinearity (\ref{alphaq})
allow the transformation $\vec{B}\to-\vec{B}$; therefore the magnetic
field vectors in  the figures shown below can be reversed.

Equation (\ref{mfd}) is nondimensionalized in terms of length $R$, the
characteristic disc radius, and the magnetic diffusion time across the
galactic disc, $h^2/\eta_0$ with $h$ the disc scale height and
$\eta_0$ a typical value of $\eta$.   Magnetic field is
measured in units of $B_0=\sqrt{4\pi\rho_0 v_{\rm t}^2/\xi}$
(corresponding to equipartition between the turbulent and magnetic
energies), where $\rho_0$ is the maximum density.  The induction
effects arising from turbulence and large-scale motions are quantified
by dimensionless dynamo numbers
\[
R_\alpha=\frac{\alpha_0 h}{\eta_0}\;,
\quad
R_\omega=\frac{\Omega_0 h^2}{\eta_0}\;,
\]
where subscript zero denotes a characteristic value of the
corresponding variable.  From here onwards we refer only to
dimensionless quantities, unless explicitly stated otherwise.

Although we are studying an $\alpha^2\omega$ dynamo characterized by
two separate dimensionless parameters $R_\alpha$ and $R_\omega$, it is
still useful  to consider the dynamo number
$D=-R_\alpha R_\omega$ as a crude measure of the dynamo intensity.
$R_\omega$ is positive here because it is defined in terms of a
typical angular velocity $\Omega_0 > 0$. In the standard asymptotic
analysis for galactic dynamos
(see Sect.~VI.4 in Ruzmaikin et al.\ 1988), $D$ is
defined in terms of $rd\omega/dr$, which is negative. Thus we have
introduced a minus sign into our definition of $D$ so that, consistent
with Ruzmaikin et al., the standard galactic dynamo has $D<0$.  Our
results are primarily a function of $D$ (for given $f_\alpha$,
$f_\eta$, see Eq.~(\ref{aleta})) being quite insensitive to the
relative values of $R_\alpha$ and $R_\omega$ in the parameter range
considered; this indicates that rotational velocity shear dominates in
the production of the azimuthal magnetic field (i.e.\ we have
approximately an $\alpha\omega$ dynamo).  The dynamo can maintain the
regular magnetic field if $|D|\geq D_\crit$, where
$D_\crit$ is a critical dynamo number, which depends weakly on details of
the model (Sect.~VI.4 in Ruzmaikin et al.\ 1988); as a rough
estimate, $D_\crit\simeq8$ in the simple model discussed in
Sect.~\ref{basicdyn}.  This marginal value (obtained with the
velocity field of A92) is close to that found for a calculation using
only the axisymmetric part of the azimuthal velocities.

The computational domain formally is $-1\leq x,y \leq +1$, but we
solve only in the region $r = (x^2+y^2)^{1/2} \leq 1$, with boundary
conditions  $B_x, B_y=0$ at $r=1$. Field amplitudes are found to be
small near the boundaries of the domain in most cases
considered,
and so the boundary conditions
remain self-consistent throughout the simulations.  The region $r>1$
is ignored; formally $\vec{B}=0$ there.  Our choice is conservative,
being arguably the least favourable for dynamo action; for example
dynamo action occurring outside $r=8\kpc$ could feed field into the
region, interior to this radius, where the equations are solved.

Our standard computational grid has $160$ mesh lines in both the
$x$- and $y$-directions, uniformly distributed over $-1 \leq x,y
\leq 1$; we made a trial integration on a finer grid, finding
insignificant differences in our results.

\subsection{Units}
\label{params}
We choose the size of the formal computational domain to be 16\kpc,
consistent with that of A92. Thus $R=8\kpc$, and we choose
$h=0.4\kpc$, although the overall nature of the solutions is
relatively insensitive to the latter choice: $h$ appears in the
dimensionless control parameters $R_\alpha$ and $R_\omega$, and the
overall field strength depends on their product $D$.
With
$\Omega_0=150\kms\kpc^{-1}$ (of order the value in the inner
$1\kpc$ radius), we have $R_\omega\approx 72/\eta_{26}$, where
$\eta_{26}=\eta_0/10^{26}\cm^2\s^{-1}$ and, if $\alpha_0$ is a few
km\,s$^{-1}$, then $R_\alpha$ will be of order unity. We assume that
$v_t=10\kms$ everywhere, so that, with $\rho_0\approx 3.5\times
10^{-24}\gcmcube$, the unit magnetic field strength is $B_0=6.6\times
10^{-6}\xi^{-1/2}\G$.

Neither $v_{\rm t}$ nor $c_{\rm s}$ are well known from observations
of barred galaxies. However, Englmaier \& Gerhard (1997) found that
sound speeds less than about $20\kms$ are necessary to produce shocks
shifted away from the bar axis.
 We note in this
connection that random velocities of molecular clouds in the central
regions of NGC~1097 are as high as $35\kms$ (Gerin et al.\ 1988), i.e.\
 3--4 times larger than the turbulent velocities at larger
distances from the centre. A similar central enhancement in the
velocity dispersion of molecular gas has been detected in NGC~3504
(Kenney et al.\ 1993).
We take $v_{\rm t}=10\kms$ as a uniform background value, but the turbulent
intensity is assumed to be enhanced in the shock fronts and in the central
region as described in Sect.~\ref{enhanc}.

A representative value of the turbulent magnetic diffusivity is
$\eta_0\simeq\frac{1}{3}v_{\rm t}l\simeq10^{26}\cm^2\s^{-1}$ for
$v_{\rm t}=10\kms$, where $l\simeq100$\,pc is the turbulent scale. A
convenient estimate of $\alpha$ is (Sect.~V.4 in Ruzmaikin et al.\
1988)
\begin{eqnarray}        \label{alph}
\alpha&\simeq&\min{\left(\displaystyle\frac{l^2\Omega}{h},v_{\rm t}\right)}\\
&=&\min{\left[0.5\kms\,\left(\displaystyle\frac{\Omega}{20\kms\kpc^{-1}}\right),
10\kms\right]},
\end{eqnarray}
where we neglect possible local enhancements of turbulent velocity beyond
$10\kms$.
With the above values of $l$, $h$ and $\Omega_0$, we obtain
$\alpha_0\simeq4\kms$. Thus, our representative values of the dynamo
coefficients are $R_\alpha=5$ and $R_\omega=70$. Since Eq.~(\ref{alph}) is just
an order of magnitude estimate, we consider that a variation of a factor 3 in
the dynamo coefficients is quite acceptable.

\subsection{Enhancement of turbulence by shear}
\label{enhanc}

\begin{figure}
\centerline{\includegraphics[width=0.99\hsize]{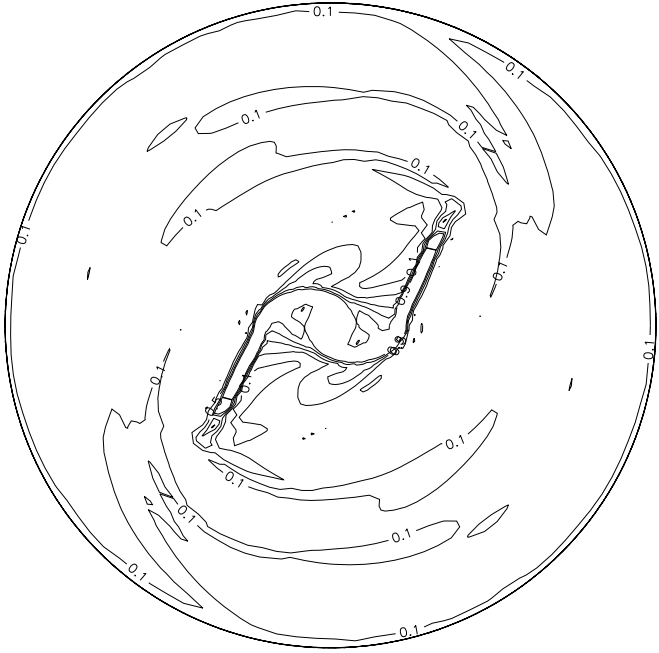}}
\caption{\label{sheareta}Contours of $\eta/\eta_0-1$ with $f_\eta=3$ in Eq.~(\protect\ref{aleta}) with
$\vec u$ from A92, as used in Sect.~\protect\ref{sheardyn}.
The 5 contour levels are equally spaced with a range from 0.10  to 0.90.
As in Fig.~\ref{vel0}, the frame radius is 1.7 times the corotation radius.
} \end{figure}

We allow for the possibility that both $\eta$ and $\alpha$ may be
enhanced where shear flow instabilities are likely to occur, by
writing
\begin{equation}                            \label{aleta}
\alpha=\alpha_0\left(1 + f_\alpha\frac{S}{S_{\rm max}}\right),\quad
\eta=\eta_0(\vec{r})\left(1 + f_\eta \frac{S}{S_{\rm max}}\right),
\end{equation}
where $f_\alpha$ and $f_\eta$ are constants and $S=|\partial
u_x/\partial y| + |\partial u_y/\partial x|$, $\vect{u}=(u_x,u_y)$
and $S_{\rm max}$ is the maximum value of $S$.  For given values of
$R_\alpha$, $R_\omega$ and/or $D$, the local dynamo action is
weakened if $f_\eta > 0$, and enhanced if $f_\alpha > 0$.  Contours
of $\eta/\eta_0-1$ with $f_\eta=3$ are shown in Fig.~\ref{sheareta};
our models have magnetic diffusivity enhanced in the dust lanes and
in the innermost 1--2\,kpc.

The effects of varying $\alpha$ and $\eta$ cannot be disentangled
completely since
it is the ratio $\alpha/\eta^2$ that affects magnetic fields
generated by an $\alpha\omega$ dynamo.  Therefore, the result of
enhanced $\alpha$ can be reproduced, quite closely, by reducing
$\eta$ instead. Thus, in order to keep the models as simple as
possible, we preferred to take $f_\alpha=0$ in most simulations.
Furthermore, with $S$ defined after Eq.~(\ref{aleta}), the turbulent
transport coefficients $\alpha$ and $\eta$ with
$f_\alpha,f_\eta\neq0$ would be enhanced even in rigidly rotating
regions. The most important result of this is that $\eta$ can be
unrealistically large in the central parts of the disc in models
discussed below. A perhaps more physically meaningful prescription
could be $S=|\partial u_x/\partial y+\omega| + |\partial u_y/\partial
x-\omega|$.  However, the effect of the latter refinement is reduced
when working in the rotating frame and, moreover, it would plausibly
be similar to that of reducing $f_\eta$.  So we leave this refinement
for future, more advanced models.  A trial calculation showed that
the effect is small when $f_\eta\la 3$ and $\alpha_0$ is constant.

We suggest that the enhancement of the turbulent diffusivity in the
shear flow (the dust lane) is due to the development of flow
instabilities.  The enhancement will be sufficient to reduce the
contrast in the magnetic field if the turbulent diffusion time is
shorter than the advection time across the dust lane width: $d^2/\eta
< d/c_{\rm s}$ with $d\simeq0.5\kpc$ in NGC~1097 (Regan et al.\ 1995,
1997) and we have assumed that the transverse velocity immediately
behind the shock is equal to the speed of sound $c_{\rm s}
\simeq10\kms$ (e.g.\ Roberts et al.\ 1979, Englmaier \& Gerhard
1997).  This yields $\eta>1.5\times10^{27}\cm^2\s^{-1}$ for the
turbulent magnetic diffusivity within the dust lanes, so $\eta$ would
have to be increased about tenfold from the background value
$\eta_0$, in order to reduce the field contrast.

The Kelvin--Helmholtz instability is an obvious candidate for
turbulence amplification in the dust lanes (see Townsend (1976 ) or
Terry (2000) for a recent discussion of turbulence in shear flows
and Ryu et al.\ (2000) and Br\"uggen \& Hillebrandt (2001) for a discussion of
mixing enhancement by the instability).
The growth rate of the instability
{\bf in a piecewise continuous flow}
is given by
\[
\gamma_{\rm KH}=\frac{U}{2d} K
\left[2K^{-1}-1-2K^{-2}\frac{\sinh{K}}{\exp{K}}\right]^{1/2},
\]
where $K=kd$ with $k$ the wavenumber in the direction of the shear
flow, $d\simeq0.5\kpc$ is the transverse width of the sheared region
and $U\simeq100$ kms$^{-1}$ is the velocity difference across it. The
growth rate has a maximum, $\gamma_{\rm max}\approx0.1U/d$, for
$K\approx0.8$. The resulting enhancement in the turbulent velocity can
be estimated as $v_{\rm t}\simeq v_{\rm t0}\exp(\gamma_{\rm
  max}d/c_{\rm s})$, where $v_{\rm t0}\simeq c_{\rm s}\simeq10\kms$
is the turbulent velocity upstream of the dust lane,
and we have assumed that the transverse velocity immediately after
the shock is equal to the speed of sound, so
the residence time of the matter in the unstable region is given by
$d/c_{\rm s}$. This yields $\gamma_{\rm max}d/c_{\rm s}\simeq1$, so
the turbulent velocity in the dust lane is expected to be 2--3 times
larger than the upstream value of $10\kms$.  The most unstable mode
has a scale of $l=2\pi d/0.8\approx4\kpc$.  The resulting
enhancement in the turbulent diffusivity
$\eta\simeq\frac{1}{3}v_{\rm t}l$
can be as large as by a factor of
100, but this enhancement will be limited by the fact
that only a part of the flow produced by the instability will be
randomized, and simultaneously the effective value of $l$ will be
reduced.
Arguably, a better estimate of $\eta$ resulting from the instability
is $\eta\simeq\frac{1}{3}v_{\rm t}^2\gamma_{\rm max}^{-1}\simeq3v_{\rm t0}d
\simeq15v_{\rm t0}l$, implying enhancement by a factor of 50. The above estimates
are upper limits, referring to the maximum rather than the mean enhancement.
The enhancement
of $\eta$
used in our models ranges up to a factor of
$f_\eta+1=6$, and the case $f_\eta=3$ is illustrated in
Fig.~\ref{sheareta}.

A magnetic field parallel to the shear velocity can suppress the
Kelvin--Helmholtz instability if $V_{\rm A}>U$ (Chandrasekhar 1981),
but this inequality is not satisfied in our case.

{\bf Apart from the Kelvin--Helmholtz instablity, enhanced small-scale
three-dimensional motions can arise from local perturbations to the
gravitational field as discussed by Otmianowska-Mazur et al.\ (2001).}

\begin{figure}
  \centering
  \newlength{\explwidth}
  \setlength{\explwidth}{0.97\hsize}
  \includegraphics[width=\explwidth]{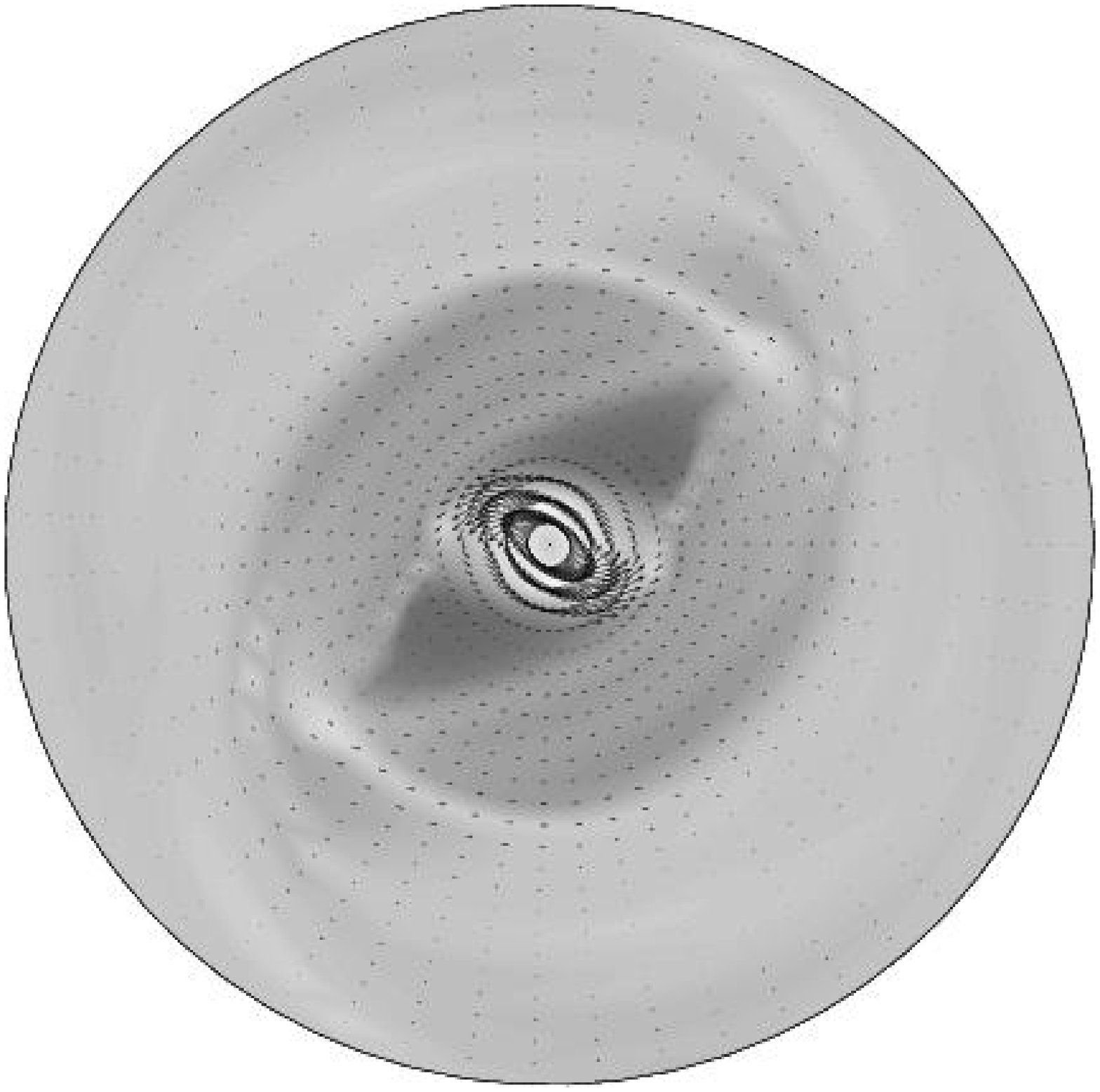}\\[5pt] 
  \includegraphics[width=\explwidth]{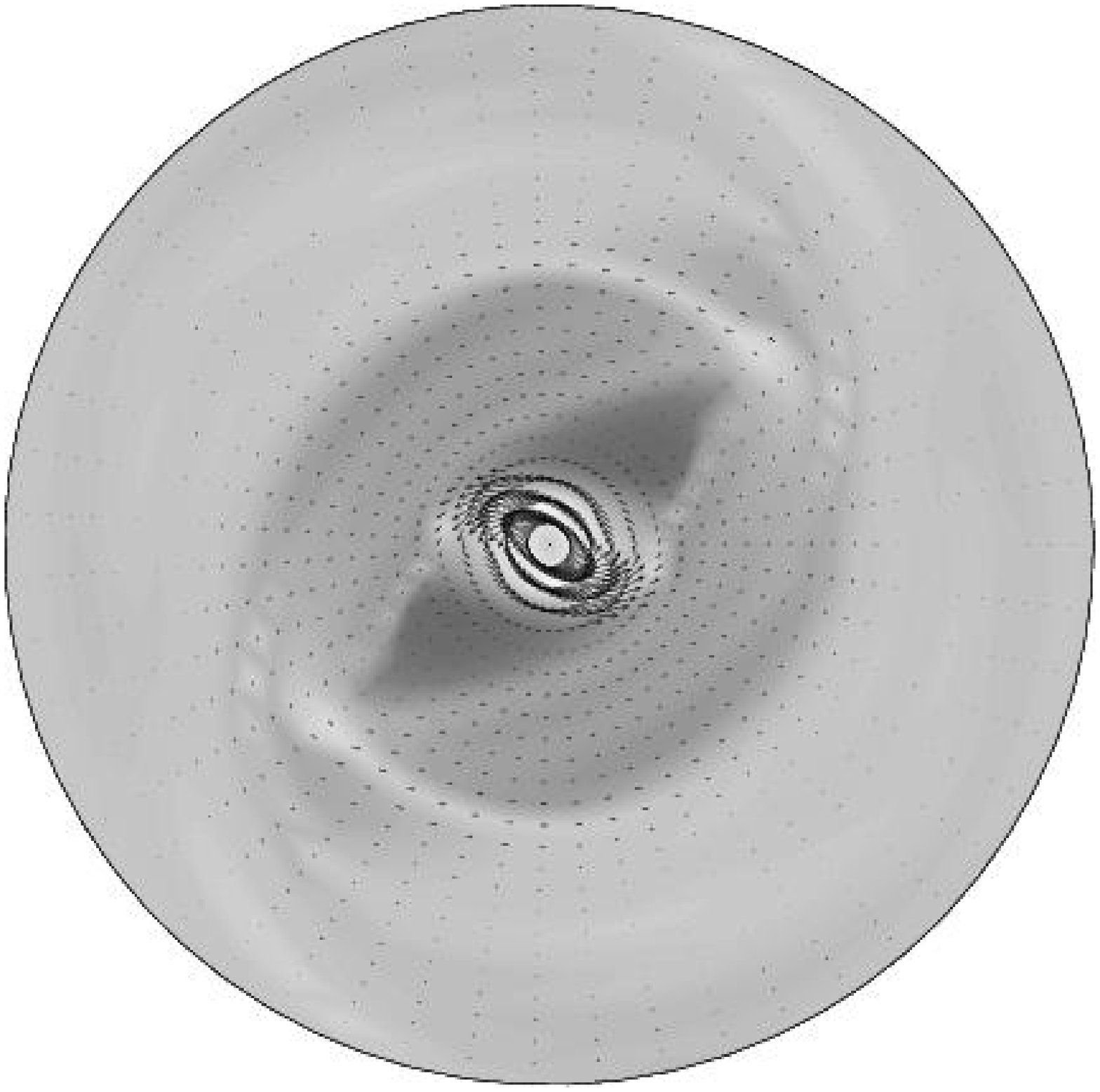}\\
  \makebox[\explwidth][l]{\raisebox{2\explwidth}[0pt][0pt]{\ \large{(a)}}}\\
  \makebox[\explwidth][l]{\raisebox{\explwidth}[0pt][0pt]{\ \large{(b)}}}\\
  \vspace*{-1\baselineskip}
  \caption{\label{model101}Vectors of the horizontal magnetic field for the
model with $R_\alpha=0.3$, $R_\omega=36$, $f_\alpha=f_\eta=0$. The
circumscribing circle has radius 8\kpc. Shades of grey show gas density, with
lighter shades corresponding to higher values. The magnetic field vectors have
been rescaled by $\vec{B}\to\vec{B}/B^n$ with $n=0.3$, to improve the
visibility of the field vectors in regions with small $B$. Panel (a) shows the
field vectors over the entire computational domain, whereas in (b) the inner
region has been omitted, in order to show the structure in the outer regions of
the disc.
}
\end{figure}

\section{Results}
\label{mainres}
\subsection{Our basic model }
\label{basicdyn}
We first examined the simplest dynamo model with
$\widetilde\alpha(\vec{r})=1$, $f_\alpha = f_\eta =0$, i.e.\ $\eta$ is
constant, unmodified by the shear, and $\alpha$ is
a function only of
the magnetic field
and gas density, via the $\alpha$-quenching.  We experimented
with several parameter combinations, and first discuss here a model
with $\eta_0=2\times 10^{26}\cm^2\s^{-1}$ so $R_\omega= 36$. With
$R_\alpha=0.3$ (giving $D=-10.8$), the dynamo is slightly
supercritical
in the central parts of the galaxy ($r<1$--2\,kpc) and subcritical
at larger radii.  The eventual magnetic field configuration is
steady in the corotating frame, and its structure is shown in
Fig.~\ref{model101}, superimposed on the gas density.  Here, and in
the other figures, we have smoothed the magnetic field over a scale
of about 0.6\kpc, to approximate the resolution of the observations.
For comparison, the radio observations have a beamwidth of
$10\arcsec$ ($0.8\kpc$) for NGC1097 and
$15\arcsec$ (2\kpc) for NGC~1365.  We see that the magnetic field is
strongly concentrated to the central regions
($r\la2\kpc$ in this
model with uniform diffusivity, reflecting the strong dynamo action
in this region where the angular velocity and its shear are maximal.
This feature remains even when the dynamo number is very substantially
increased.
In this model, the maximum field strength
of about $0.5B_0$ is reached at $r\approx0.7\kpc$ (corresponding to
about $3\mkG$); the field is negligible beyond $r\simeq2\kpc$.
Observations of NGC~1097
suggest that the ratio of the regular field strength in the inner
region ($r/r_{\rm c}\simeq0.1$) to that in the outer bar region
is about 2, and so this model is unsatisfactory.
Note that this feature is
little altered by the smoothing to the resolution of the
observations.

Even when $|D|$ is increased by an order of magnitude, the
model magnetic field
remains strongly concentrated to the central regions.
Note that in all our models the field strength increases with $|D|$,
but in order to obtain field strengths comparable with those observed
near to
the corotation radius, it is necessary to increase $|D|$ to values
at the margin of what
is plausible, and the field in the inner regions is then significantly
larger than observed.
The overall
field structure is relatively insensitive to the magnitude of $D$.
An obvious problem with this basic model with $\eta=\mbox{const}$ is
that it yields an unrealistically weak magnetic field in the
outer regions and perhaps a too strong contrast in $B$ between the dust
lane regions and those upstream of them.
Thus we now relax the condition that $\eta$ be constant.

\begin{figure*}[!htb]
  \newlength{\exprwidth}
  \setlength{\exprwidth}{0.49\textwidth}
\centerline{
        \includegraphics[width=\exprwidth]{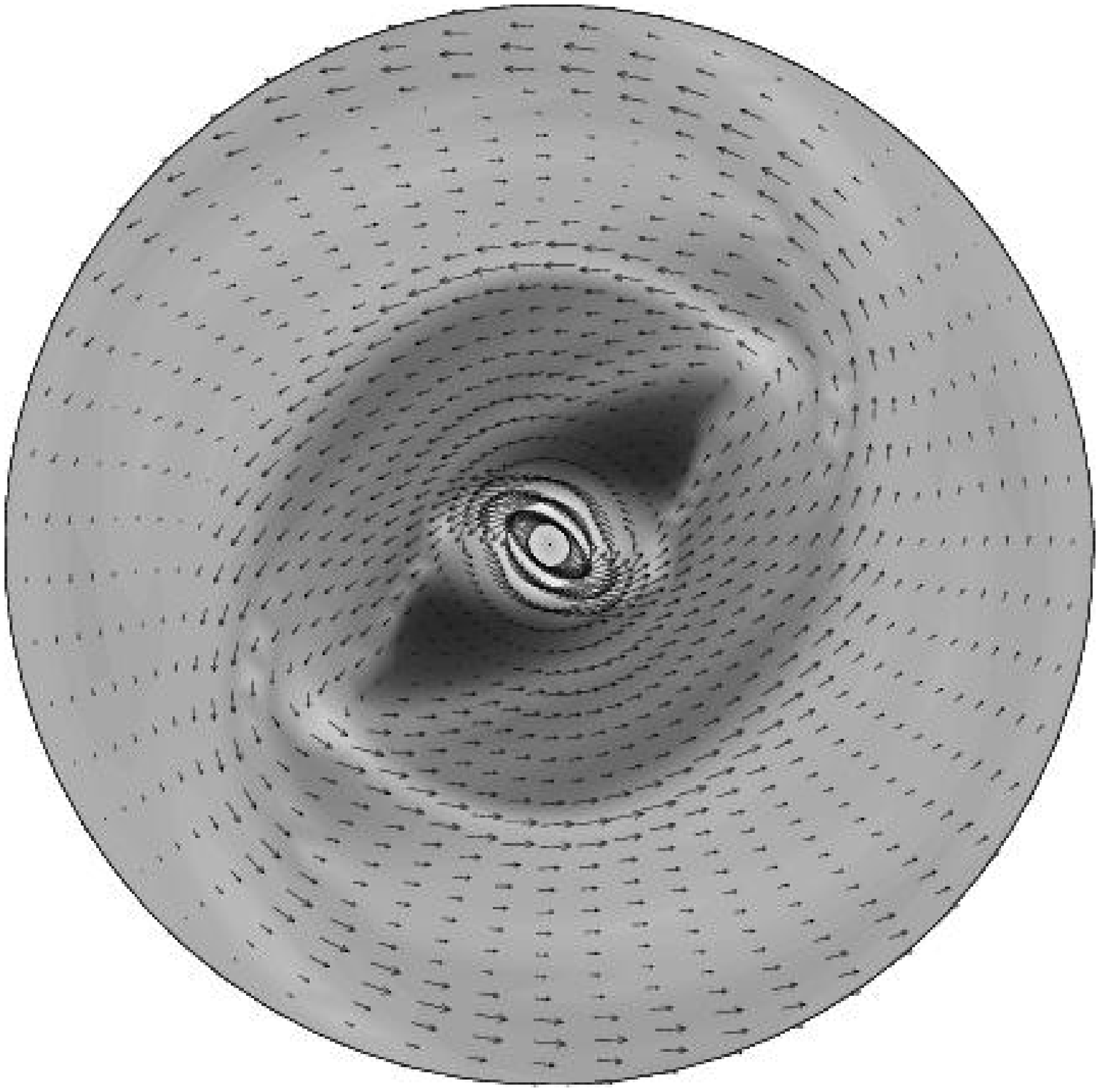}
  \hfill
        \includegraphics[width=\exprwidth]{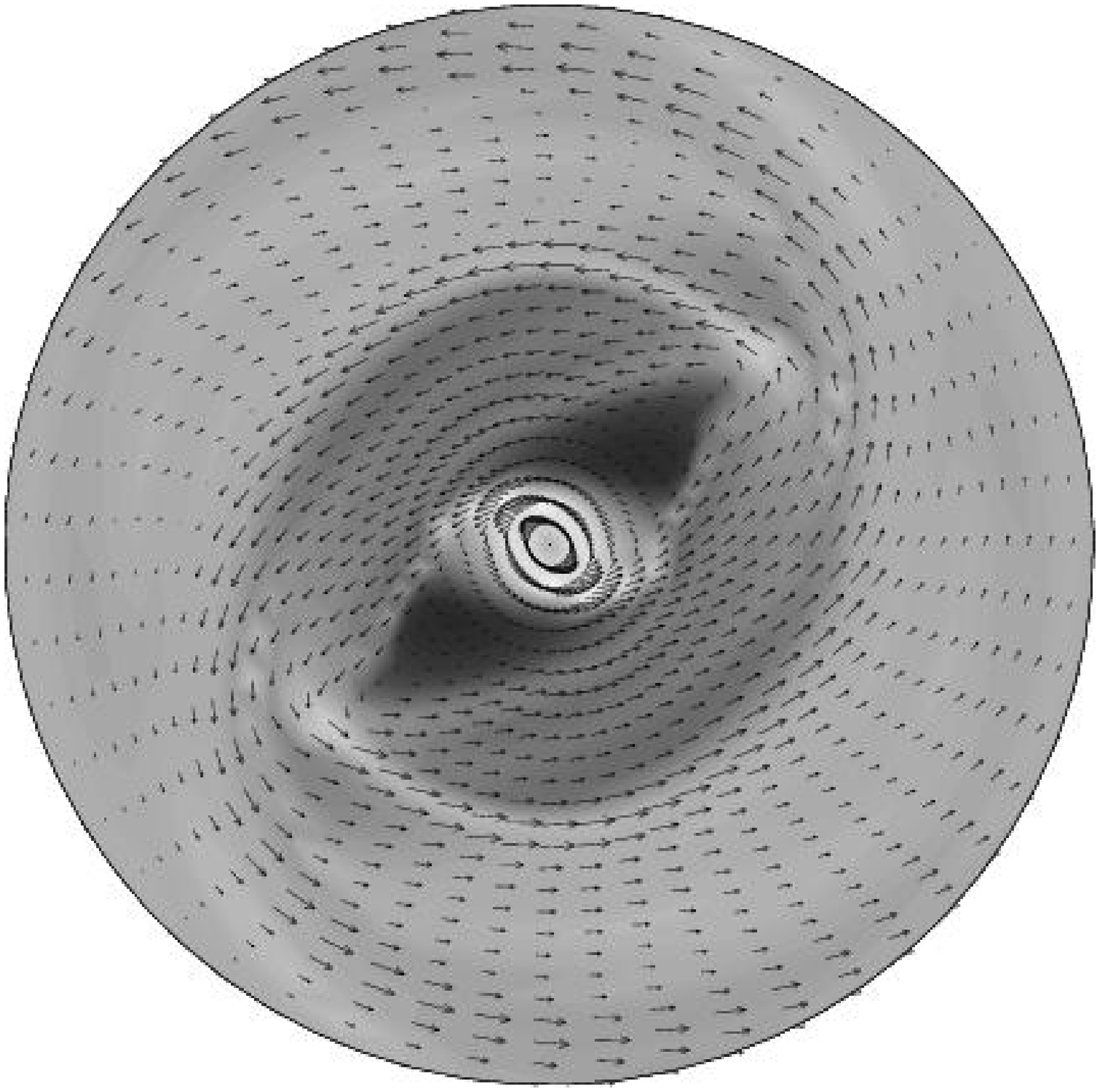}}
\vspace*{5pt}
\centerline{
        \includegraphics[width=\exprwidth]{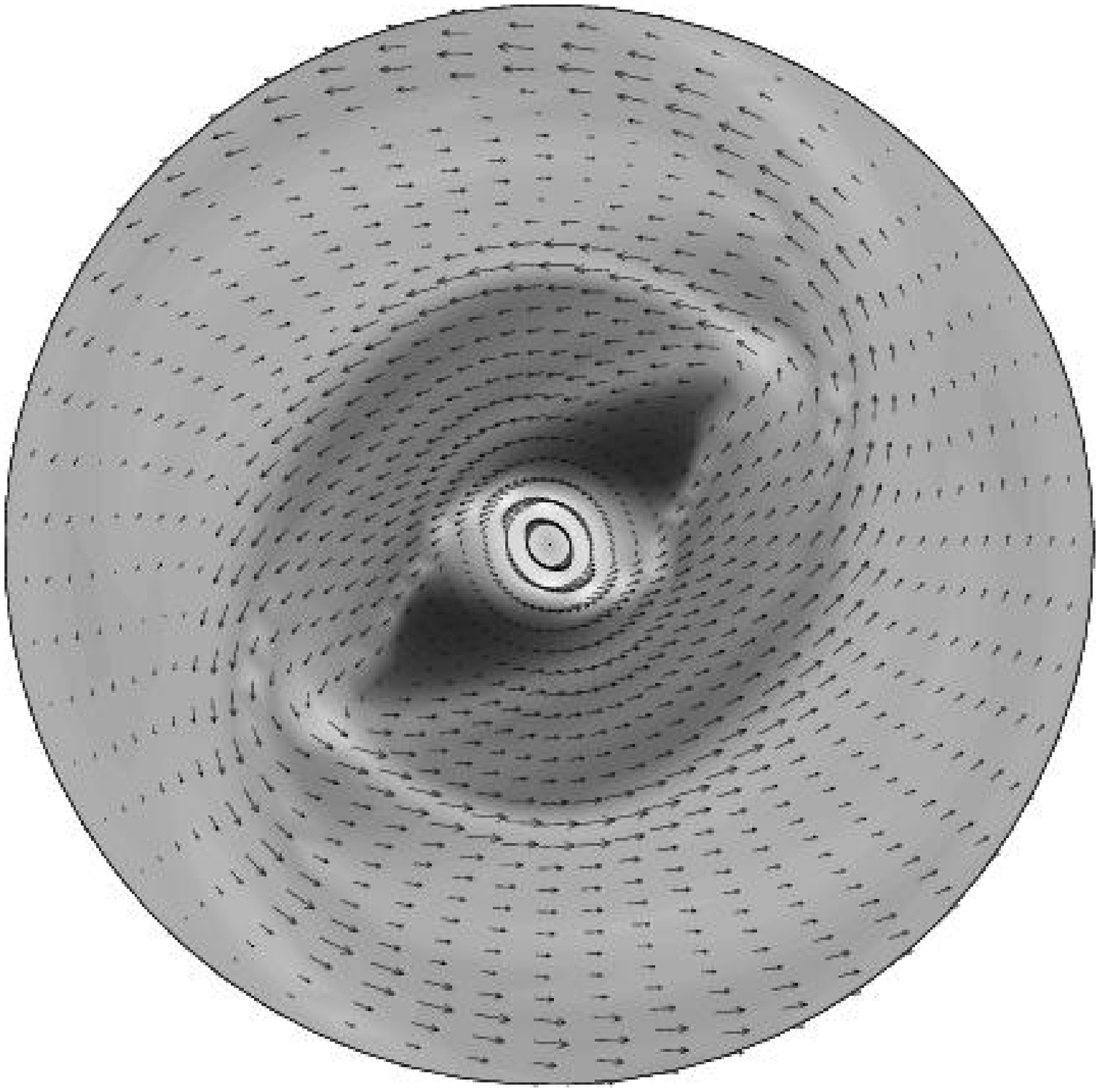}
  \hfill
        \parbox[b]{\exprwidth}{
\caption{\label{modelunif}Vectors of the horizontal magnetic field for the
models with  $\widetilde\alpha(\vec{r})=1$, $f_\alpha=0$, $R_\alpha=3$, $R_\omega=72$.
(a) $f_\eta=1$,  (b) $f_\eta=3$,  and (c) $f_\eta=5$.
The circumscribing circle has radius 8\kpc. Shades of grey show gas density, with
lighter shades corresponding to higher values.
The magnetic field vectors
have been rescaled as $\vec{B}\to\vec{B}/B^n$ with $n=0.3$, to improve the
visibility of the field vectors in regions with small $B$.}}}
  \makebox[\exprwidth][l]{\raisebox{2\exprwidth}[0pt][0pt]{\ \large{(a)}}}\hfill%
  \makebox[\exprwidth][l]{\raisebox{2\exprwidth}[0pt][0pt]{\ \large{(b)}}}\\
  \makebox[\exprwidth][l]{\raisebox{\exprwidth}[0pt][0pt]{\ \large{(c)}}}\\
  \vspace*{-2\baselineskip}
\end{figure*}

\subsection{Models with turbulent diffusivity depending on velocity
  shear: uniform $\widetilde\alpha$} \label{sheardyn}
An obvious refinement of the model is to consider the turbulent
magnetic diffusivity to be modulated by the shear rate, i.e.\ to put
$f_\eta>0$ in Eq.~(\ref{aleta}), keeping the background value uniform,
i.e.\ $\eta_0(\vec{r}) = \mbox{const}$.  Note that, as $f_\eta$ and thus the
magnetic field dissipation increase, so does the marginal dynamo
number, for example $D_\crit\approx -27$ when $f_\eta=5$, and even
$D=-11$ is just subcritical when $f_\eta=1$.
Also, broadly speaking the effects of taking $f_\alpha>0$ for given
$R_\alpha$ and of increasing $R_\alpha$ with $f_\alpha=0$ are similar,
and so we keep $f_\alpha=0$.
We now set $\eta_0=10^{26}$ cm$^2$s$^{-1}$, so that $R_\omega=72$ and
adopt $R_\alpha=3$.
Thus the models we discuss now
are all substantially supercritical, with the
local values of $D$ (calculated from the local values of $\Omega$)
of order $-(10^2$--$10^3)$ at $r\la3\kpc$ and
about $-15$ near the corotation radius.

In Fig.~\ref{modelunif} we show the field structures for calculations
with $f_\eta = 1, 3, 5$.  Again, the magnetic fields are steady in
the rotating frame.  Increasing $f_\eta$ clearly reduces the
dominance of the central field. Also, increasing $\eta_0$ to $2\times
10^{26}$ cm$^2$s$^{-1}$ ($R_\omega=36$) produces little overall
effects on the field structure,
but does affects the field strength.  When, for example, $R_\omega=36$,
$f_\eta=1$, $R_\alpha=0.6$ ($D = -22.5$), the maximum field strength
is about $0.7 B_0$; this figure increases to about $6.5 B_0$ when
$R_\omega=72$, $f_\eta=1$, $R_\alpha=3$ ($D = -216$,
Fig.~\ref{modelunif}a)
Of this increase, a factor of about 4 can
be attributed to enhanced induction by rotational shear
which yields $B\simeq B_0\left(|D|/D_\crit-1\right)^{1/2}$, and the
rest to the noncircular velocities.

A feature of all these solutions is that there is a broad field
minimum
in the bar region, but with strong magnetic
ridges at the positions of the dust lanes and field enhancement in
the central part where the circumnuclear ring is observed. This
agrees well with the overall distribution of polarized intensity in
NGC~1097 and  1365. Nevertheless, the
local magnetic energy density near the major axis (upstream of the
dust lanes) exceeds the turbulent kinetic energy density.
As we discuss in
Sect.~\ref{ShockR}, both the model and observed field strengths have
a well pronounced structure within the bar region in addition to the
ridges elongated with the dust lanes and the central ring.

\begin{figure}
\centerline{\includegraphics[width=0.99\hsize]{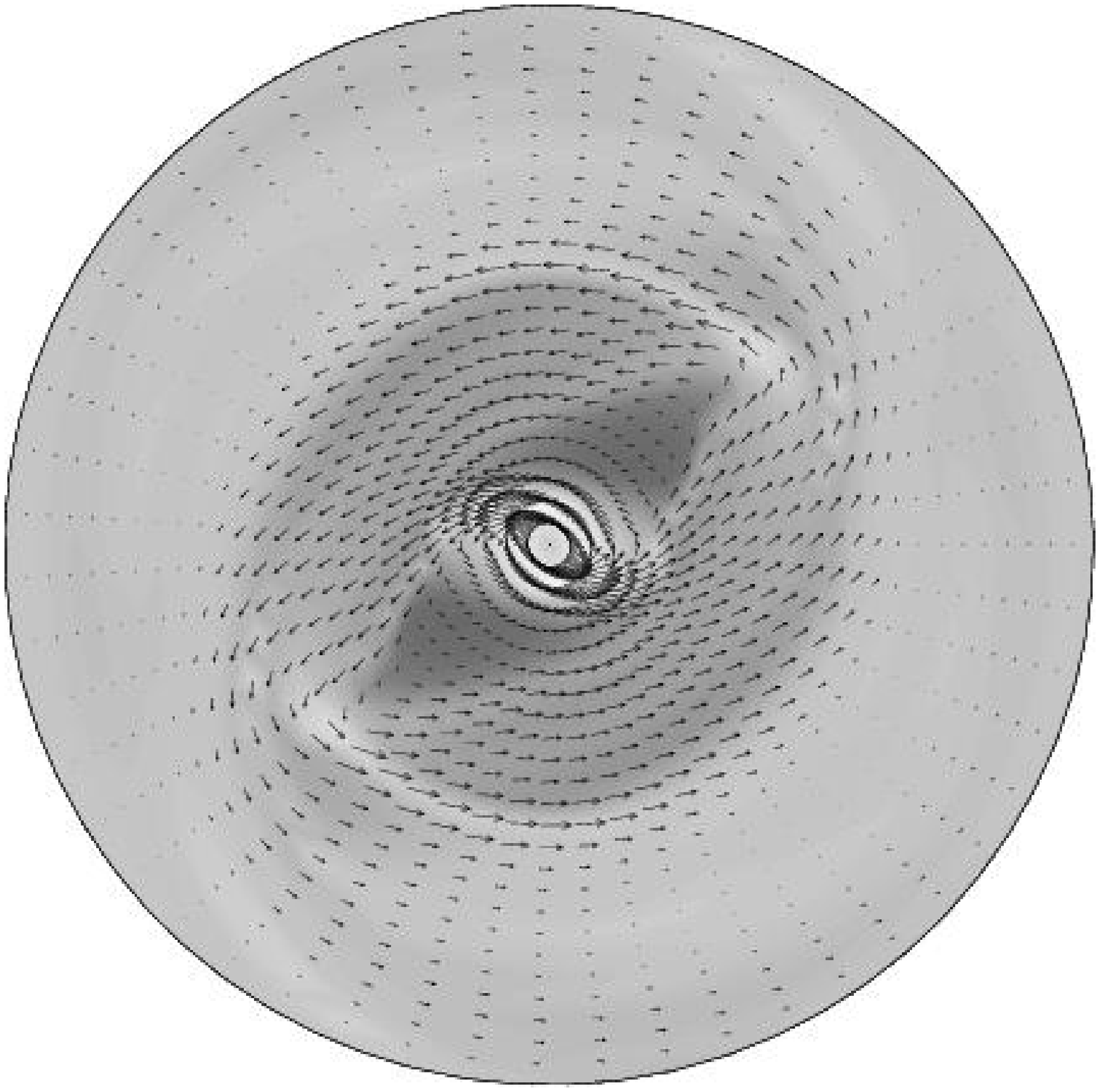}}
\caption{\label{Dsmall}As in Fig.~\ref{modelunif}a, but for
$R_\alpha=0.6$, $R_\omega=36$, $f_\eta=1$,
giving weak dynamo action.
}
\end{figure}

There is a relatively strong, quasi-azimuthal field upstream
of the shock, especially near the ends of the bar. This feature also
agrees well with the observations, cf.\ Fig.~\ref{N1097obs}.
For the dynamo parameters of the models illustrated in Fig.~\ref{modelunif},
the dynamo is locally supercritical nearly everywhere.
However, the field structure is rather insensitive to the value of $D$.
To illustrate this, we
show in Fig.~\ref{Dsmall} the field
vectors with only a slightly supercritical dynamo number
($\eta_0=2\times 10^{26}\cm^2\s^{-1}$, $R_\alpha=0.6$,
$R_\omega=36$). The field configuration is remarkably similar
{\bf to that}
of Fig.~\ref{modelunif}a, with a much larger dynamo number, except
perhaps near the outer boundary.
This supports the idea that the
field in regions upstream of the dust lanes at larger
galactocentric distances is mainly produced not by local dynamo
action but rather by advection over almost $180^\circ$ from regions
at small radii (where the dynamo action is stronger) on the other
side of the bar.
We note that the half-rotation time of the gas at $r\simeq5\kpc$ is shorter than
magnetic diffusion time $h^2/\eta_0\simeq5\times10^8\yr$. If the
dynamo is strong enough ($R_\alpha\ga 3$ for $R_\omega=72$) the advection
produces a magnetic configuration where the magnetic
energy density exceeds that of the interstellar turbulence in broad
regions, including the region upstream of the shock front.  The steep
rotation curve at small radii and the large local dynamo numbers
found there suggest that
the dynamo is particularly efficient
in the inner region.  This field is then transported
outwards by the noncircular velocities and further enhanced by shear.

\begin{figure}
\includegraphics[width=0.99\hsize]{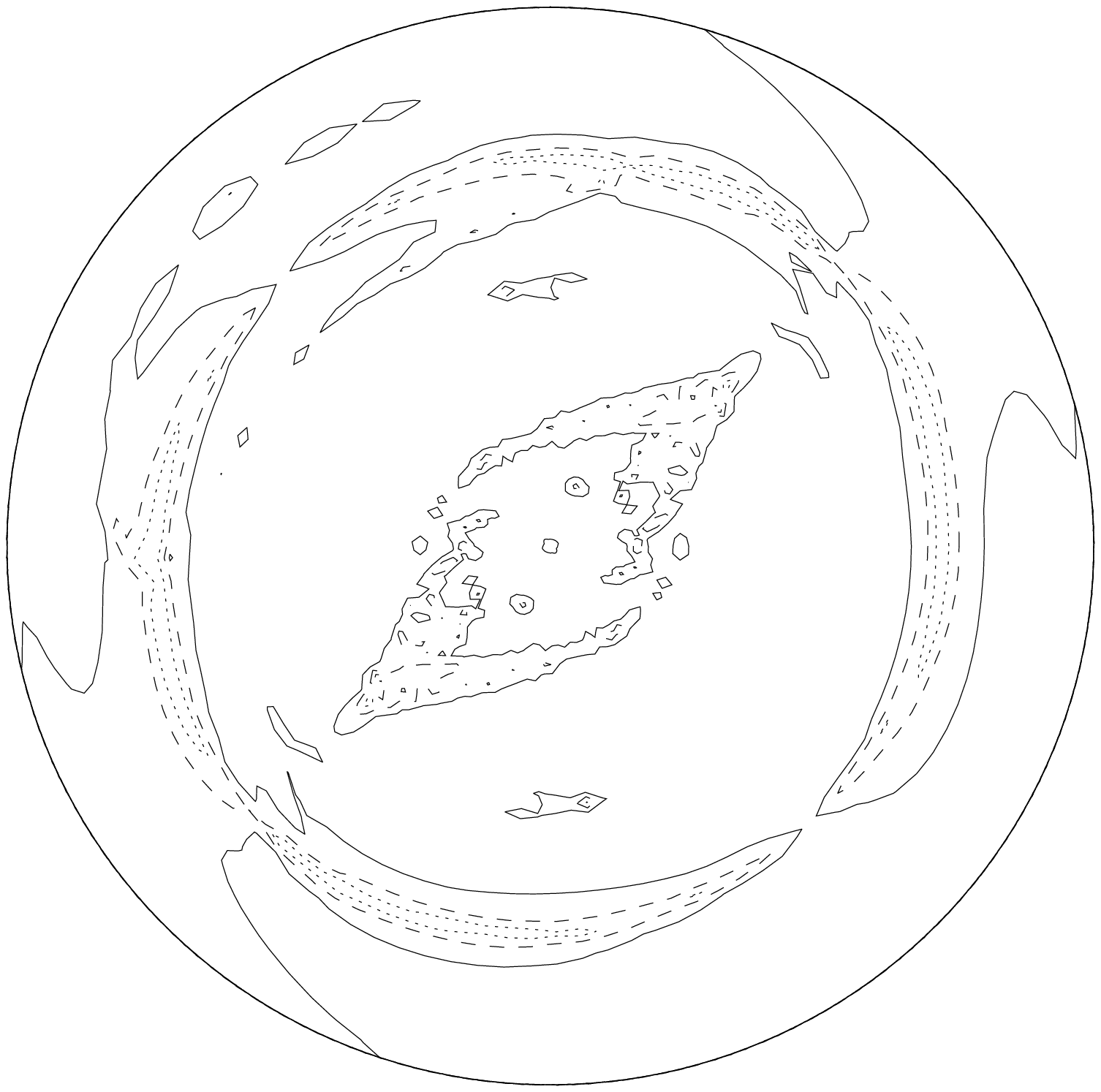}
\caption{\label{angle}The alignment of the regular
magnetic and velocity fields in the corotating frame is illustrated
with level contours of $\cos\chi=|\vec{u}\cdot\vec{B}|/uB$ for the model
of Fig.~\ref{modelunif}a. The levels shown are 0.4
(dotted), 0.7 (dashed) and 0.95 (solid).
}
\end{figure}

Immediately downstream of the shock position, the velocity and
magnetic fields are closely aligned.
The alignment is fairly good throughout the galactic
disc, with the mean angle $\chi$ between magnetic and velocity
vectors being about $\chi\approx20^\circ$
($\langle\cos\chi\rangle\approx0.93$) in the model of
Fig.~~\ref{modelunif}a.  As can be seen in Fig.~\ref{angle}, the
alignment is especially close outside the bar and slightly reduced
within the bar and at the corotation radius. The reduction in the bar
can be plausibly attributed to enhanced magnetic diffusivity in the
dust lanes and that at corotation to the small local values of
$\vec{u}$ in the rotating frame.

\begin{figure}
\centerline{\includegraphics[width=0.99\hsize]{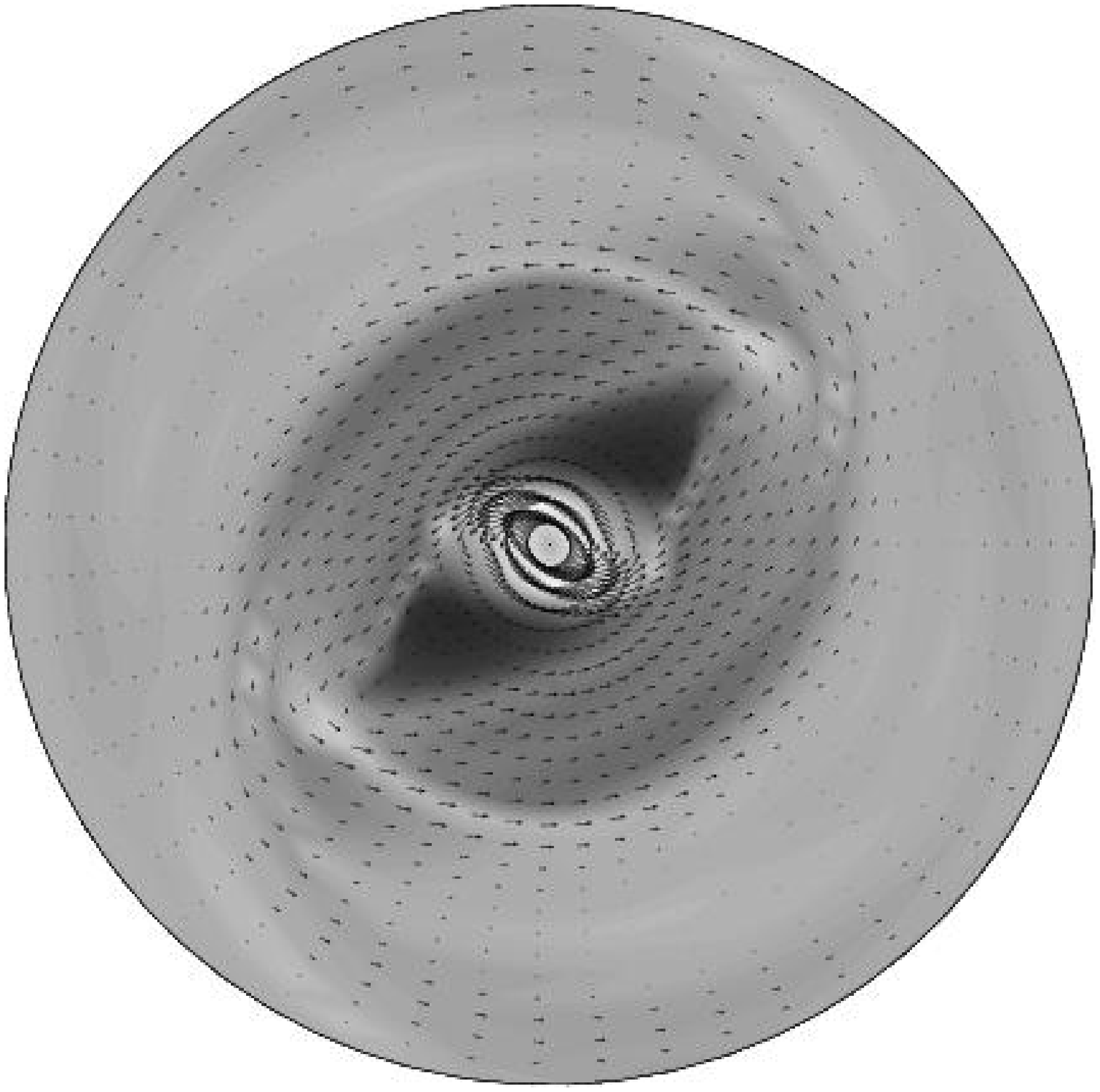}}
\caption{\label{alpprom}As in Fig.~\ref{modelunif}a, but for $\widetilde\alpha(\vec{r})\propto\omega(r)$
and $R_\alpha=1.5$.
}
\end{figure}

\begin{figure}
\centerline{\includegraphics[width=0.99\hsize]{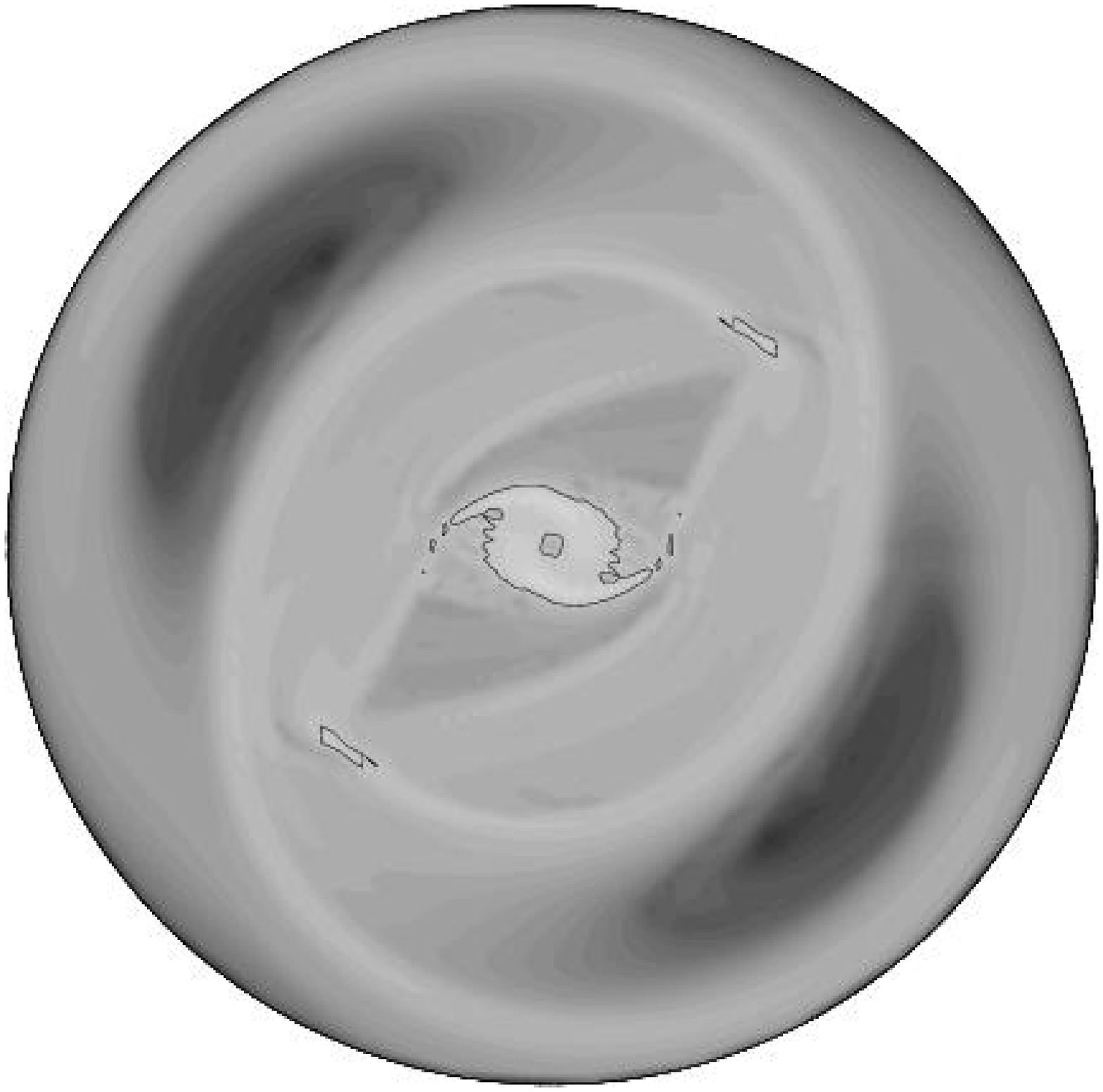}}
\caption{\label{BB}Grey scale map of $B^2$ (unsmoothed)
for the model shown in Fig.~\protect\ref{alpprom}.
Lighter shades indicate higher values,
and the {\bf contours shown are for $B^2=1$, so the energy density of the
regular magnetic field exceeds that of turbulence inside the contours.}
The circumscribing circle has radius 8\kpc.
}
\end{figure}

\subsection{Models with turbulent diffusivity depending on velocity
shear: $\widetilde\alpha \propto \omega$} \label{sheardyn1}

We now discuss models in which we set $\widetilde\alpha\propto
\omega(r)$ (see Sect.~\ref{basic}), whilst keeping $f_\alpha=0$.
Since the dynamo action now becomes more intense in the disc centre,
the field strength is more strongly concentrated to the centre for
given $f_\eta$, than when $\widetilde\alpha$ is uniform.
As $f_\eta$ is varied, these results can be broadly
summarized by saying that the field structure {\it is} very similar
to that when $\widetilde\alpha(\vec{r})$ is uniform, {\it if} a
larger value of $f_\eta$ is now taken.  For example, we show in
Fig.~\ref{alpprom} the field structure when $R_\omega=72$,
$f_\eta=3$, $R_\alpha=3$, which can be compared with that shown in
Figs.~\ref{modelunif}a and \ref{modelunif}b, with $\widetilde\alpha(\vec{r})=1$
and $f_\eta=1, 3$ respectively.
We note that the effect of setting $\widetilde\alpha
\propto \omega(r)$ is the opposite of increasing $f_\eta$ -- the
former enhances the central field concentration, the latter decreases
it. But the field structure outside of the inner 1--2\,kpc is altered
only slightly by such changes.

For this model, the maximum field strength is about $3.9 B_0$,
attained at $r/r_{\rm c}\approx0.1$ (see also Fig.~\ref{radB2}).
We show in Fig.~\ref{BB} a grey scale plot of $B^2$ for the model
of Fig.~\ref{alpprom}
where magnetic field enhancements in the dust lanes, the spiral
arms and the central region are prominent. The field is especially strong in the
circumnuclear region and at the ends of the bar.
Trailing magnetic
arms, emanating approximately from the ends of the bar, are a common feature
of the modelled and observed magnetic structures. (This feature is also seen
in the model for the weakly barred galaxy IC~4214 of Moss et al.\
1999.)

This class of models exhibits the best agreement with
observations and we discuss them in detail in Sect.~\ref{Discuss}.

\subsection{Magnetic fields in the absence of dynamo action: the case
of pure advection} \label{advect}

\begin{figure}
\centerline{\includegraphics[width=0.99\hsize]{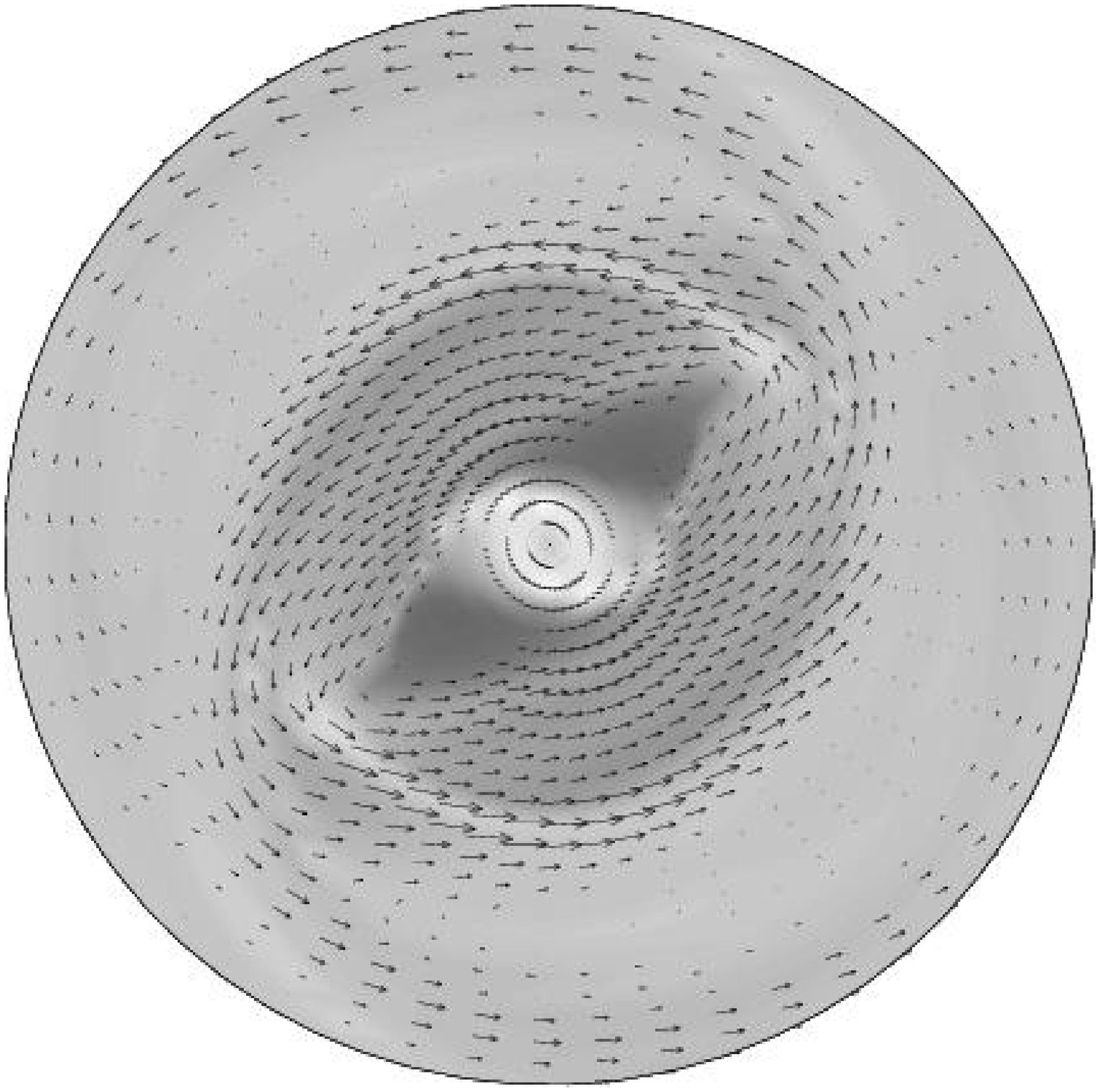}}
\caption{\label{adv}As in Fig.~\ref{modelunif}, but without any dynamo
action: $R_\alpha=0$, $R_\omega=72$,
  $f_\alpha=0$ and $f_\eta=5$.
In this case the magnetic field decays on a time scale of about $10^8\yr$.}
\end{figure}

In this section we attempt to elucidate the roles of the two key
ingredients of our model, stretching by the regular velocity and the
$\alpha$-effect. In Fig.~\ref{adv} we show the field structure for a
computation with $R_\omega=72$, $f_\eta=5$, and $R_\alpha=0$ (no
dynamo action). In the absence of dynamo action  this field decays
exponentially, with an $e$-folding time of about $10^8\yr$.  This decay
time scale is about 5 times shorter than in normal galaxies because of
the enhanced velocity shear that reduces the field scale and so
facilitates its decay.  Since most barred galaxies in the sample of
Paper I do exhibit detectable polarized radio emission and thus
possess regular magnetic fields, we conclude that the need for dynamo
action in barred galaxies is even stronger than  in
normal galaxies.

Looking at the spatial structure of the decaying magnetic field, we
see that some of the features of the dynamo maintained fields shown in
Figs.~\ref{modelunif} and \ref{alpprom} are present, particularly the
field minimum upstream of the shock, and the
trailing magnetic arms outside of the corotation radius.  However
there is no hint of the relatively strong fields at small radii seen
in
all models with dynamo action present. In addition, the relative magnetic field
strength near the bar major axis is now too low.

Thus it appears that the field structure in the outer regions is relatively
insensitive to the parameters of the dynamo model, but that dynamo action is
essential to maintain the fields over time scales of gigayears and to produce
strong magnetic fields in the region of the circumnuclear ring
and upstream of the shock front.


\subsection{Results with positive dynamo number}
\label{extra}
We made some experiments with $R_\alpha < 0$, i.e.\ positive dynamo
number $D$.  This is motivated by suggestions that this may be the
case in accretion discs (Brandenburg et al.\ 1995) even
though the origin of turbulence there is quite different from that in
galaxies.  It is a familiar result for axisymmetric galactic discs
that the marginal positive dynamo numbers for dynamo excitation are
much larger in magnitude than the marginal values for the usually
considered case with $D<0$ -- i.e.\ dynamos with $D>0$ are much harder
to excite than those with $D<0$.  As mentioned in Sect.~\ref{basic}, in
the models with $D<0$ the marginal values of $D$ for the models with
the full A92 velocity field are essentially unchanged from those found
by including only the axisymmetric rotational velocities.  When $D>0$,
the situation is quite different. A dynamo is excited by the full
velocity field at a marginal dynamo number of magnitude considerably
less than required to excite the corresponding dynamo driven by the
circular motions alone.
In other words, when $D>0$ the dynamo is driven by shear due to noncircular
velocities rather than from differential rotation.

\begin{figure}
\includegraphics[width=0.99\hsize]{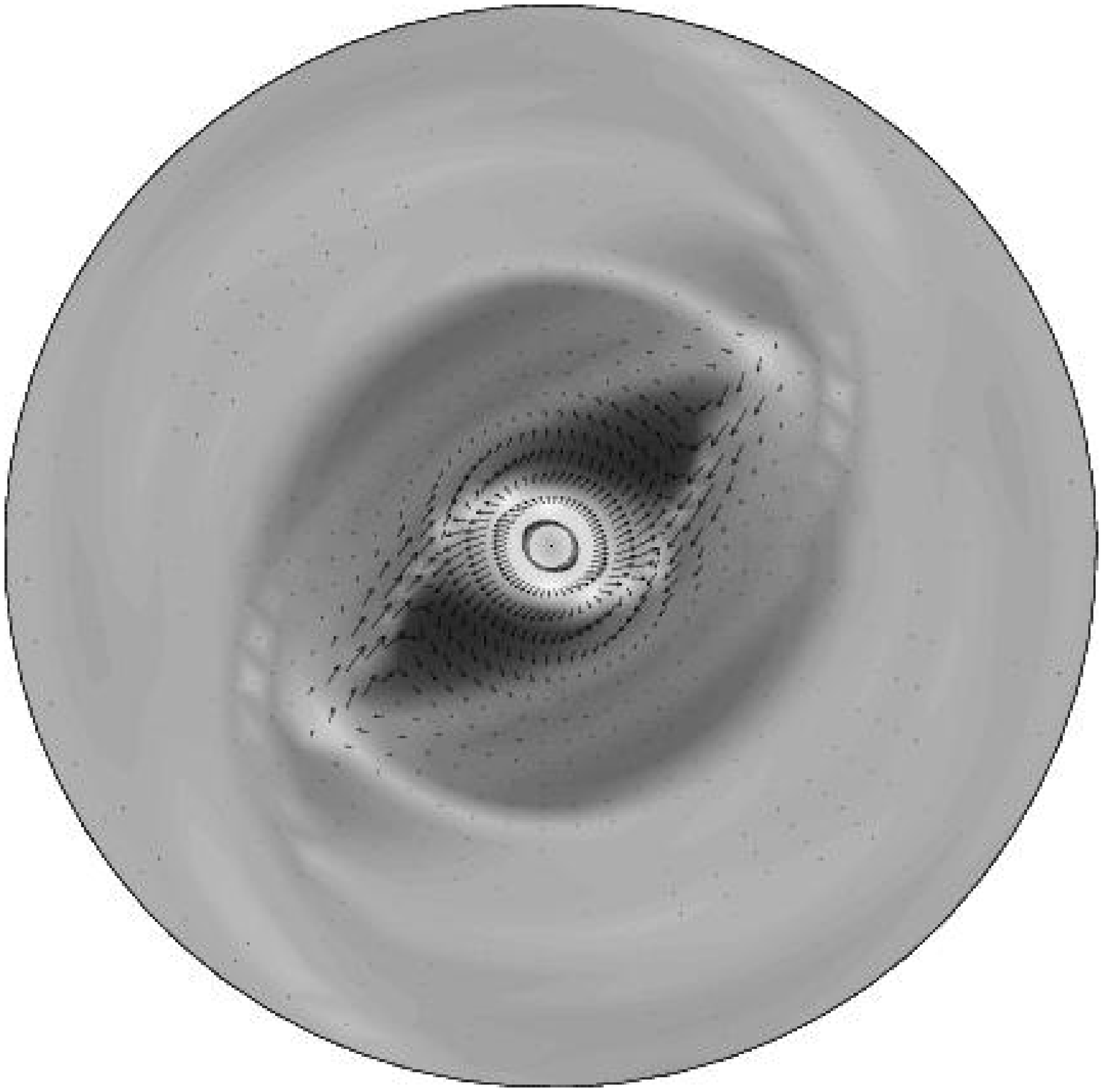}
\caption{\label{model57}Snapshot of typical vectors of the horizontal magnetic
field for the positive dynamo number model with
$R_\alpha=-6$, $R_\omega=36 $, $f_\alpha=0$, $f_\eta=5$,
in a format similar to that of Fig.~\ref{modelunif}.
}
\end{figure}

\begin{figure*}
  \newlength{\expqwidth}
  \setlength{\expqwidth}{5.5cm}
        \resizebox{1.1\expqwidth}{!}{\includegraphics{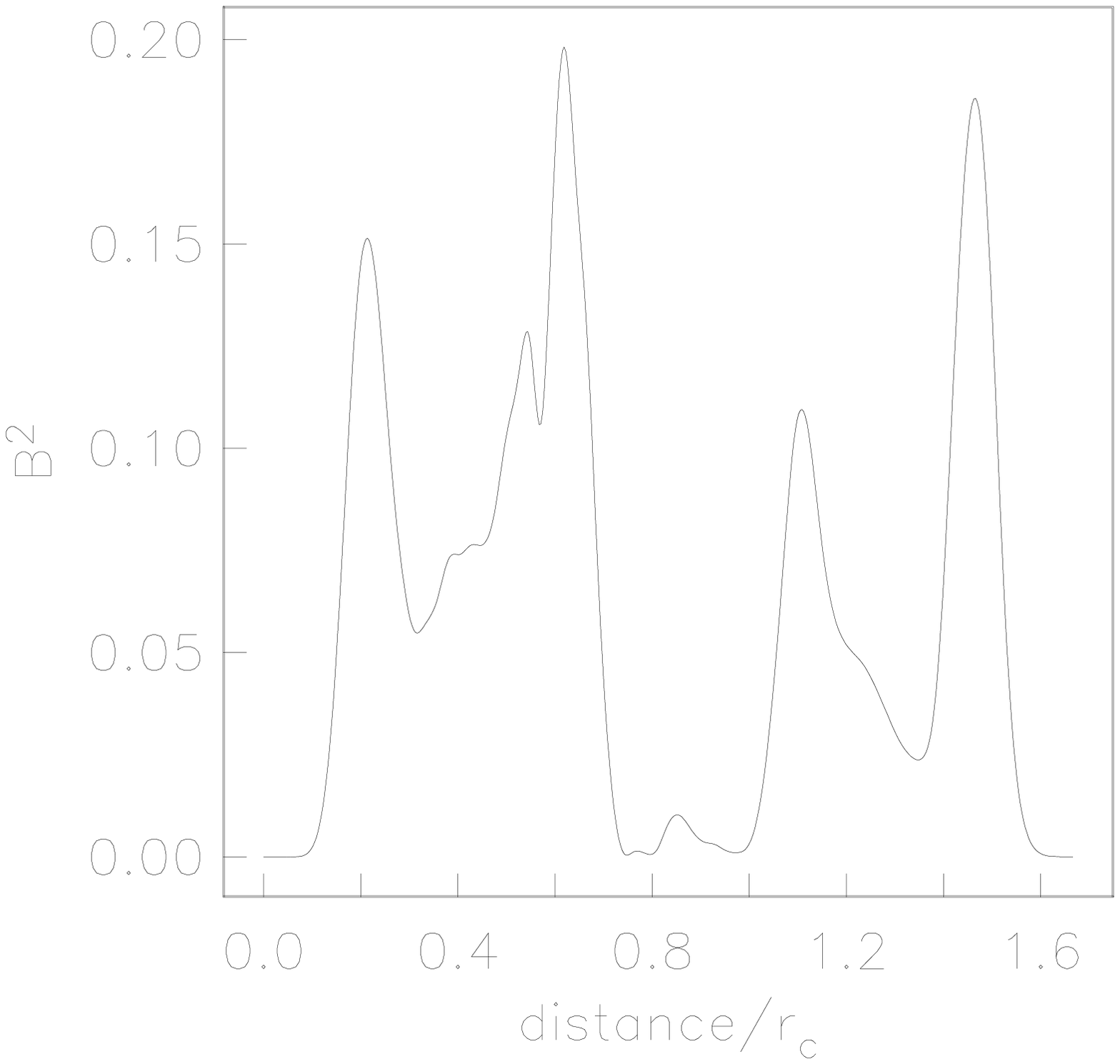}}
        \resizebox{1.1\expqwidth}{!}{\includegraphics{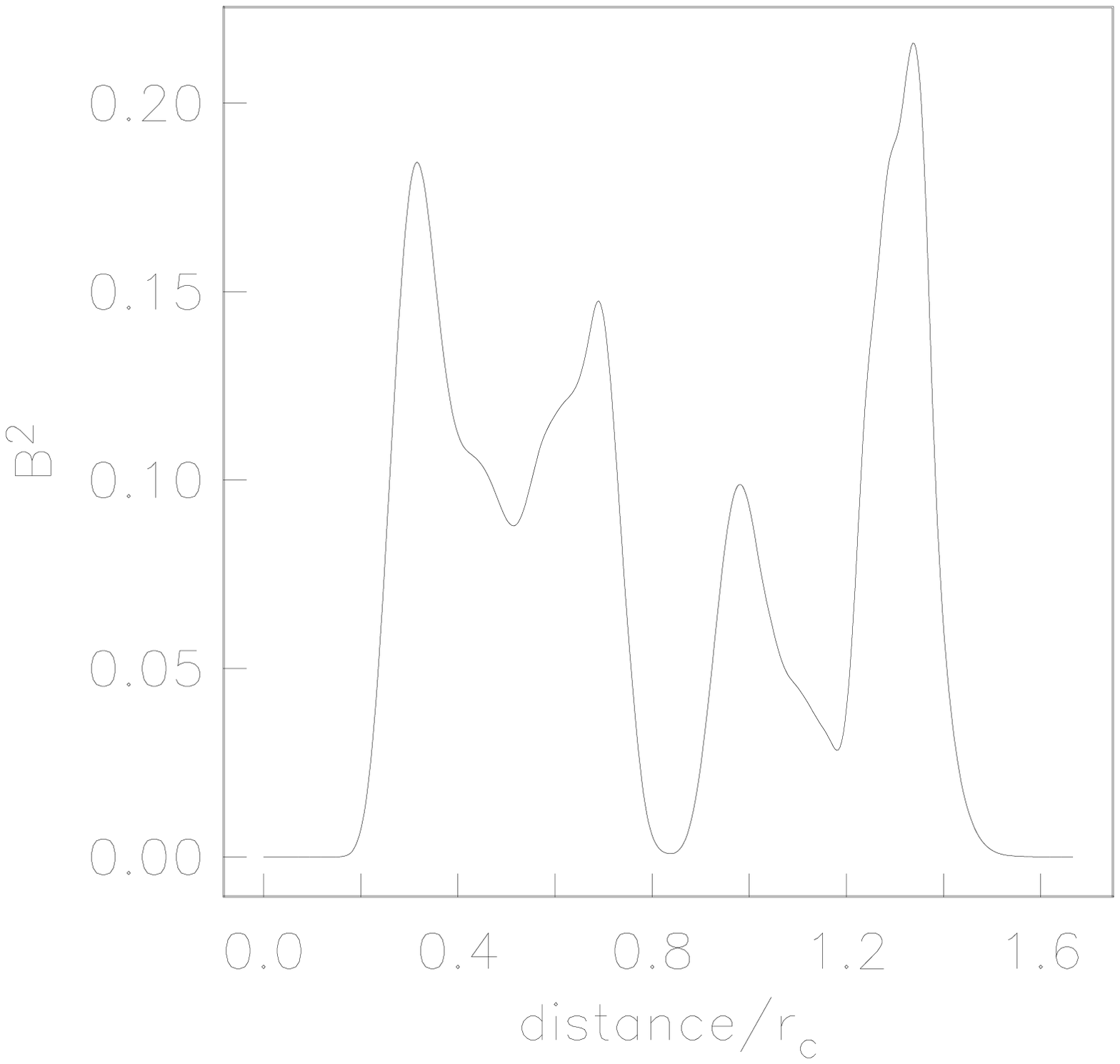}}\\[-4\baselineskip]
\hspace*{0.08\expqwidth}
        \resizebox{\expqwidth}{!}{\includegraphics{fig12c.eps}}
\hspace*{0.08\expqwidth}
        \resizebox{\expqwidth}{!}{\includegraphics{fig12d.eps}}
\hfill
\parbox[b]{55mm}{
\caption{\label{cuts}Profiles of the square of the horizontal magnetic
field along cuts perpendicular to the shock front of the model
shown in Fig.~\ref{alpprom}
at (a) 1.5\,kpc and (b) 3\,kpc galactocentric distance. The total
length of each cut is 10~kpc, approximately symmetrical about the position of the shock front,
with distance increasing in the upstream direction.
The corresponding cuts (with scaled lengths of 20\,kpc;
$20\arcsec$ corresponds to $r/r_{\rm c}\approx0.14$)
through the map of $\lambda6$~cm polarized intensity
(represented by the vector length of Fig.~\protect\ref{N1097obs})
are shown in (c) and (d) for galactocentric distances of 3 and
6\,kpc. The cuts are shown for the same values of $r/r_{\rm c}$ in
the model and NGC~1097 and extend over the same linear length.}}
  \makebox[\expqwidth][l]{\raisebox{2.1\expqwidth}[0pt][0pt]{\hspace*{0.25\expqwidth} \large{(a)}}}
\hspace*{0.08\expqwidth}
  \makebox[\expqwidth][l]{\raisebox{2.1\expqwidth}[0pt][0pt]{\hspace*{0.25\expqwidth} \large{(b)}}}\\
  \makebox[\expqwidth][l]{\raisebox{1.05\expqwidth}[0pt][0pt]{\hspace*{0.25\expqwidth} \large{(c)}}}
\hspace*{0.08\expqwidth}
  \makebox[\expqwidth][l]{\raisebox{1.05\expqwidth}[0pt][0pt]{\hspace*{0.25\expqwidth} \large{(d)}}}
  \vspace*{-2\baselineskip}
\end{figure*}

We illustrate our results by discussing a simulation with
$R_\omega=72$, $f_\eta=5$, $f_\alpha=0$, $\widetilde\alpha(\vec{r})=1$
and $R_\alpha=-6$, a somewhat supercritical value. The magnetic
fields are now oscillatory, with a dimensionless period of about
0.48, corresponding to about $1.1\times 10^8$ yr.  We see in
Fig.~\ref{model57} that the regions of strong magnetic field
correspond to the regions of strong velocity shear.  The overall
field structure varies significantly during the cycle period, and at
certain times the amount of structure in the magnetic field is much
greater than at others.  The magnetic field strength has a maximum in
the inner bar region and in two elongated features parallel to the
shock fronts.  These structures persist through nearly all of the
oscillation period.  However, the field geometry within the
structures changes with time as a magnetic field reversal along a
line approximately parallel to the shock front propagates clockwise
in Fig.~\ref{model57}. In the lower left-hand part of
Fig.~\ref{model57}, weak field directed approximately radially
outwards is visible. At a slightly later time, this has grown in
strength and this region eats in to the adjacent area of
approximately inwardly directed field, strengthening the field
reversal. A little later still, the reversal has disappeared, and then
the cycle repeats.  We emphasize that, for this calculation the field
decays when the nonaxisymmetric velocities are removed.

It is clear that these dynamos with positive dynamo numbers generate
magnetic fields that are rather different from those observed.


\section{Discussion and comparison with observations}\label{Discuss}

\subsection{The magnetic field in the shock region} \label{ShockR}
We show in Fig.~\ref{cuts} the variation of $B^2$ along two cuts
through the shock front region for the model illustrated in
Fig.~\ref{alpprom}, illustrating the variation in the
magnetic field strength between the post- and
pre-shock regions.  Also shown is the observed profile of polarized
synchrotron intensity along  similarly chosen
cuts in NGC~1097. In
the case of constant energy density of cosmic-ray electrons, the
polarized intensity depends on
$B_\perp^2$, where $B_\perp$ is the magnetic field component
perpendicular to the line of sight.

Both the model magnetic field and the observed polarized
intensity  reveal several peaks in $B^2$
on each side of the shock front and a pronounced minimum at the
shock front itself.  The model and observations differ in detail,
which is no surprise as {\it inter alia\/} the velocity field and gas
distribution used in our models was not designed specifically to fit
NGC~1097. We find that the best correspondence between the $B^2$ variations from
the models and the observed polarized intensity profiles is obtained
in models with $\widetilde{\alpha}\propto \omega$ (Sect.~\ref{sheardyn1}), with
$R_\alpha\la 1.5$ when $R_\omega=72$. Larger values of $|D|$ give weaker peaks
near the bar, but taking too small a value of $|D|$ gives rather low
field strengths.

Due to the inclination of
NGC~1097 ($45^\circ$), the difference between $B_\perp^2$ and
$B^2$ is typically a factor of 2. Projection effects reduce  the polarized
intensity upstream of the shock from where magnetic field has
significant component parallel to the line of sight, but not in the
dust lane where the value of $B_\perp$ is closer to that of $B$.
This will suppress the amplitude of the secondary peak in polarized intensity
in the pre-shock region. With allowance for the unaccounted projection effect,
our model shows remarkably good agreement with observations in the number and
width of the magnetic peaks in the shock region.

{\bf Our models reproduce a smooth turn of the magnetic field vector upstream
of the dust lanes observed in NGC~1097 (and also NGC~1365). This turn may be
difficult to see in the figures presented here because we have shown magnetic
vectors only at a fraction of mesh points. It can be seen from Fig.~\ref{angle}
that the alignment between the magnetic and velocity fields is reduced 1--2\,kpc
upstream of the shock fronts (dust lanes). Several effects apparently
contribute to smear and displace the turn in the magnetic field with respect to
that in the velocity field. We argue in Sect.~\ref{sheardyn} that magnetic
field can be advected to the pre-shock region from smaller radii (the elongated
shape of the streamlines responsible for that can be clearly seen in Fig.~2b of
A92, but not that easily in Fig.~\ref{vel0} here). The admixture of
magnetic field generated elsewhere must smear any sharp structures produced
locally. Furthermore, magnetic field will tend to be aligned with the prinicpal axis
of the rate of strain tensor rather than the velocity field itself, and so its turn can
start long before it becomes visible in the velocity vectors.}

{\bf Other effects can be envisaged that might, in principle, result in a smooth turn of
magnetic field upstream of the shock fronts. For example, this can be a
manifestation of the dynamical influence of the regular magnetic field on the
velocity field. However, this is hardly the case in our models where the regular
magnetic field does not reach equipartition with even the turbulent kinetic
energy in the regions concerned (see Fig.~\ref{BB}). Another
possibility is that the turn in the observed field is smoothed by the presence of
a magnetic halo with a regular magnetic field smoother than in
the disc; however, there is so far no evidence for a polarized halo in NGC~1097.
A further option is that the magnetic field is anchored in a warm component
of the interstellar medium that has different kinematics than the cold gas
(i.e., higher speed of sound);
our preliminary analysis of this possibility has given no evidence for this,
but we shall return to it elsewhere.}

\begin{figure}
\centerline{\includegraphics[width=0.99\hsize]{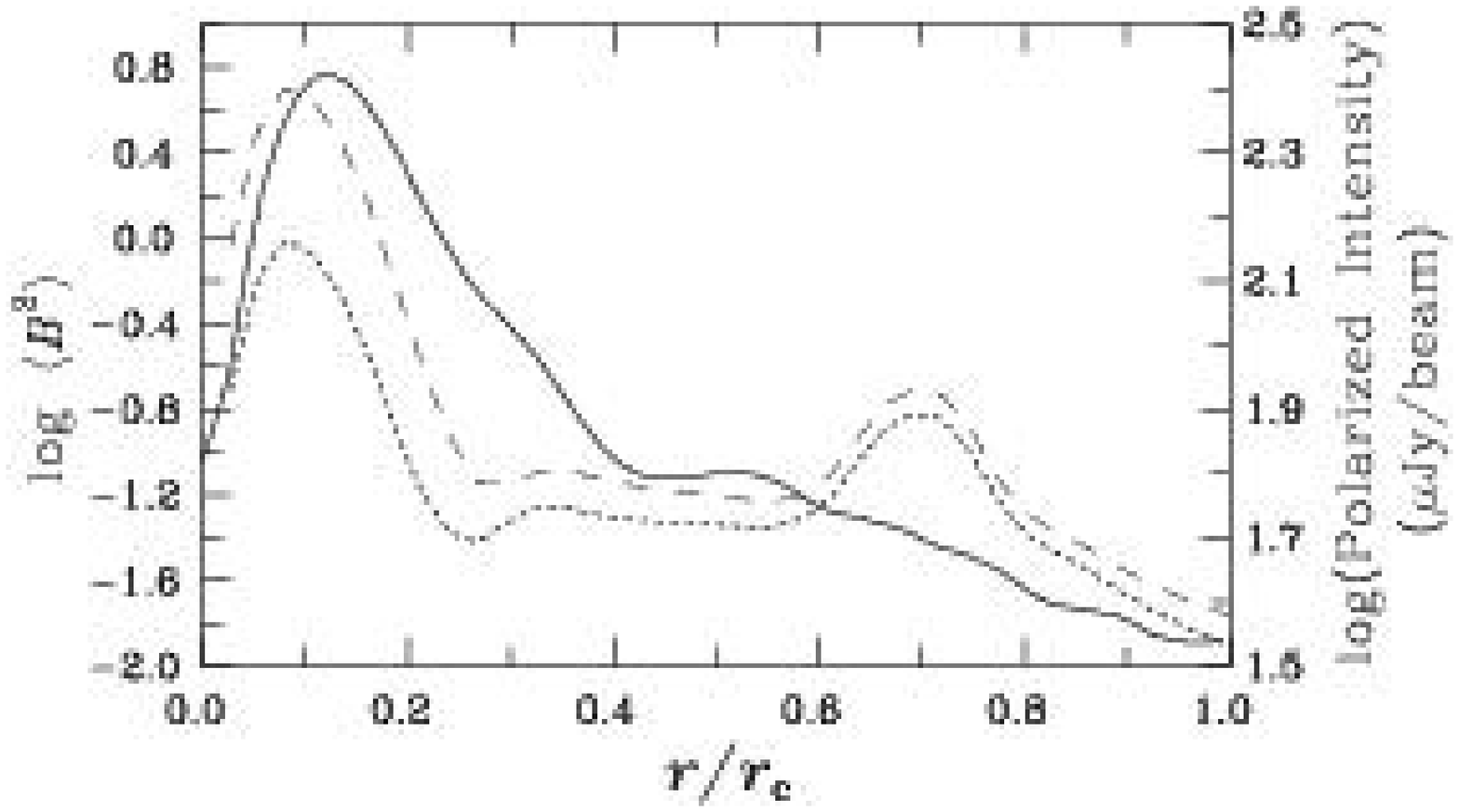}}
\caption{\label{radB2}The radial dependence of the azimuthally averaged square
of the horizontal magnetic field for the models with $R_\alpha=1.5,\ R_\omega=
72,\ \widetilde{\alpha}\propto\omega$, but with $f_\eta=1$ (dashed)
and $f_\eta=3$ (dotted). For comparison the radial profile of the azimuthally
averaged polarized intensity at $\lambda6\cm$ in NGC~1097 is shown with
solid line. {\bf Note the logarithmic scale on the vertical axes;
$\langle B^2\rangle=0$ at $r=0$ in our models.}
 }
\end{figure}

\subsection{Radial profiles of magnetic field}
We show in Fig.~\ref{radB2} the dependence of the azimuthally
averaged value of $B^2$ on radius for the model shown in
Fig.~\ref{alpprom}
and a similar model with larger $f_\eta$, and compare it with the observed dependence of
polarized intensity on radius.  The model reproduces the peak observed close to the
centre
and has a realistic radial scale. The slower decrease of polarized intensity
with galactocentric radius in NGC~1097 can be explained by a steeper rotation
curve in the A92 model, which has the turnover radius at $r/r_{\rm
c}\approx0.25$ whereas it occurs at about $r/r_{\rm c}=0.4$ in NGC~1097
(Ondrechen \& van der Hulst 1989). The model magnetic field has a pronounced
subsidiary peak at $r/r_{\rm c}\approx0.7$; this peak is related to the strong
magnetic arms visible in Fig.~\ref{BB}. A similar peak in the observed radial
profile is less pronounced and occurs at a smaller fractional radius.

{\bf In our models, the regular magnetic field vanishes on the rotation axis
because of its symmetry properties. However, the
observed polarized intensity remains significant at $r=0$. We discuss
implications of this in the next section.}

\begin{table}
\caption{The global magnetic structure in NGC 1097
represented in terms of the amplitudes of the azimuthal Fourier
modes ${\cal R}_m$, their pitch angles $p_m$ and phases $\beta_m$
with $m$ the wavenumber.
The bottom line shows azimuthally averaged magnetic field strengths
obtained assuming $\nel=0.1\cmcube$ and $h=0.4\kpc$.
}\label{Mmodes}
\begin{center}
\begin{tabular}{lllll} \hline \noalign{\smallskip}
  & & \multicolumn{3}{c}{Radial range (kpc)}  \\
  &  & 2.9--3.7 & 3.7--4.5 & 4.5--5.3 \\
\hline
\noalign{\medskip}
${\cal R}_0$ & $\rm{rad\,m^{-2}}$ &
  $-152$\,\scriptsize{$\pm6$} & $-141$\,\scriptsize{$\pm13$} &
  $-152$\,\scriptsize{$\pm17$} \\ \noalign{\smallskip}
$p_0$ & degrees &
  $18$\,\scriptsize{$\pm1$} & $27$\,\scriptsize{$\pm2$} &
  $22$\,\scriptsize{$\pm2$} \\ \noalign{\smallskip}
${\cal R}_1$ &  $\rm{rad\,m^{-2}}$ &
  $-75$\,\scriptsize{$\pm8$} & $-88$\,\scriptsize{$\pm12$} &
  $-90$\,\scriptsize{$\pm10$}  \\ \noalign{\smallskip}
$p_1$ & degrees &
  $-50$\,\scriptsize{$\pm13$} & $-71$\,\scriptsize{$\pm13$} &
  $-58$\,\scriptsize{$\pm7$}\\ \noalign{\smallskip}
$\beta_1$ & degrees &
  $-119$\,\scriptsize{$\pm6$} & $-98$\,\scriptsize{$\pm6$} &
  $-122$\,\scriptsize{$\pm8$}  \\ \noalign{\medskip}
${\cal R}_2$ &  $\rm{rad\,m^{-2}}$ &
  $110$\,\scriptsize{$^{+3}_{-185}$} & $57$\,\scriptsize{$^{+8}_{-145}$} &
  $75$\,\scriptsize{$^{+15}_{-8}$} \\ \noalign{\smallskip}
$p_2$ & degrees &
  $0$\,\scriptsize{$^{+1}_{-130}$} & $-7$\,\scriptsize{$^{+9}_{-102}$} &
  $-31$\,\scriptsize{$^{+8}_{-91}$} \\ \noalign{\smallskip}
$\beta_2$ & degrees &
  $13$\,\scriptsize{$\pm4$} & $11$\,\scriptsize{$\pm15$} &
  $28$\,\scriptsize{$\pm8$} \\ \noalign{\medskip}
\hline\noalign{\smallskip}
$\langle B\rangle$ &  $\mu\rm{G}$ &
  $5.3$\,\scriptsize{$^{+0.3}_{-1}$} &
        $4.9\,$\scriptsize{$\pm0.6$} &
          $5.3$\,\scriptsize{$^{+0.6}_{-1}$}\\
\noalign{\smallskip}
\hline
\end{tabular}
\end{center}
\end{table}

\subsection{The azimuthal structure of the global
  magnetic field} \label{modal}
The azimuthal structure of the regular magnetic field is a useful
diagnostic of both the overall structure of the gas velocity and the
field generation mechanisms. We have fitted the polarization angles
observed in NGC~1097 at $\lambda\lambda3.5$ and 6.2\cm, for three
radial rings between 2.9 and $5.3\kpc$ with a model where the
magnetic field is expanded into azimuthal Fourier modes. The model
and the fitting procedure are discussed by Berkhuijsen et al.\ (1997)
and Fletcher et al.\ (2001);
Table~\ref{Mmodes} gives details of the fits.  The polarization
angles have been obtained from Stokes parameters averaged over
$20^\circ$ sectors in the $0.8\kpc$ wide rings.  The results are
presented in terms of ${\cal R}_m=0.81\,B_m\nel h$, $p_m$ and
$\beta_m$, where $B_m$
(in $\mu$G)  is the amplitude of the magnetic mode
 with azimuthal wave number $m$, $\nel$ is the thermal
electron density in cm$^{-3}$, $h$ is the scale height of the thermal
ionized layer in pc, $p_m$ is the pitch angle of the mode $m$ (i.e.
the angle between the field direction and the circumference;
$p>0$ indicates a trailing spiral) and
$\beta_m$ is the azimuthal angle, measured counter-clockwise from the
northern end of the major axis of the galaxy, where the mode $m$
($m\neq0$) has maximum strength.  ${\cal R}_m$ is related to the
Faraday rotation measure; ${\cal R}_m=100\radm$ corresponds to a
field strength of about $3\mkG$ for $n_{\rm e}=0.1\cmcube$ and
$h=400$\,pc. Note that the limited azimuthal resolution of the fits
$(m\leq2)$ does not allow the fitting of the sharp deflection of the
$B$ vectors in the shock wave region (Fig.~\ref{N1097obs}).

The  amplitudes of the three lowest azimuthal Fourier modes
in the observed regular field of NGC~1097 have ratios $1:1:0.5$.
The corresponding values for the models shown in Figs.~\ref{modelunif}a,b,c
and Fig.~\ref{alpprom} are all $1:0:0.5$.
The axisymmetric mode is dominant in both the observed field and in
all these models. The relative amplitude of the $m=2$ mode
for these models is also in agreement with the observations,
having about half the amplitude of the $m=0$ mode.

The magnetic fields in our models do not contain the $m=1$ magnetic
mode present in the observed field. This is  due  to the strict
symmetry of the model velocity field (it does not contain any modes
with odd $m$) and the inability of the dynamo to maintain the $m=1$
mode on its own.  The only mode actively maintained by the dynamo in
our models is the axisymmetric one, $m=0$, and the $m=2$ mode arises
via distortion of the basic field by a flow with a strong $m=2$
component.  In turn, other magnetic modes with $m$ even are also
maintained.  In real galaxies such a high degree of symmetry of the
velocities cannot be expected, and  thus
a wider range
of azimuthal modes will be  present in the magnetic field.
We note
that NGC~1097 is perturbed by a companion, so the real velocity field
will certainly  be more
complicated than that used in the dynamo simulations.  Any $m=1$
component of velocity will then generate a corresponding component of
magnetic field from the $m=0$ field by the $\vec{u}\times\vec{B}$ term
in the induction equation (e.g.\ Moss 1995),
although whether the amplitude of the resulting $m=1$ mode is large enough to be consistent
with the observed value can only be determined by a detailed calculation.
Naive numerical experimentation suggests that the amplitude of the $m=1$ velocity component
would have to be comparable with that of the $m=2$ component if a $m=1$
magnetic field of the
strength implied by Table~\ref{Mmodes} is to be produced
solely by the  $\vec{u}\times\vec{B}$ interaction.
{\bf Anisotropy in the $\alpha$-effect, neglected here, can also excite the
$m=1$ magnetic mode via dynamo action (Rohde \& Elstner 1998).}
Neglecting the $m=1$
mode, the implication is that all the models presented in
Figs.~\ref{modelunif} and \ref{alpprom} reproduce the global
azimuthal structure sufficiently well.

The absence of modes with odd $m$ explains why
the model magnetic field
vanishes at the disc centre as all modes with even $m$ have $B=0$ at
$r=0$ from symmetry considerations. Models with a more realistic
velocity field can have strong magnetic fields with $m=1,3,\ldots$ at
the disc centre.
The
{\bf significant observed} polarized intensity at $r=0$ (see Fig.~\ref{radB2})
{\bf may be} due to magnetic modes with odd $m$.

The estimates of azimuthally averaged magnetic field
strength resulting from the fits of Table~\ref{Mmodes}, shown in
the bottom line of the Table, have been obtained using an electron
density of $\nel=0.1\cmcube$ and an ionized disc scale height of
$h=0.4\kpc$.
(The errors given do not include any uncertainty in $\nel$ and $h$.)
Both parameters are poorly known for barred galaxies.
The diffuse H$_\alpha$ flux in barred galaxies is only
moderately lower than in normal spirals (Rozas et al.\ 2000) and the
\HI\ scale height is comparable to (or somewhat larger than) in the
Milky Way (Ryder et al.\ 1995), so electron densities can be expected to
be similar in different galaxy types. The electron density
at comparable fractional radii (with respect to corotation) of 4--5\,kpc in the
Milky Way are about $0.1\cmcube$ (Taylor \& Cordes 1993). This justifies
our crude estimate $\nel=0.1\cmcube$ for $r\simeq5\kpc$ in NGC~1097.
A significantly lower value of $\nel$ would imply
a systematically larger magnetic field strength than estimates resulting from
energy equipartition with cosmic rays.

\subsection{The field in the central ring}
\label{central}
In the top right of Fig.~\ref{N1097obs}, the field at a galactocentric
radius of about 1.5--$2\kpc$ can be seen to have a strongly
nonaxisymmetric structure apparently dominated by the $m=2$ mode with
a significant $m=0$ mode also present.
The overall field structure features
strong magnetic fields at those azimuthal angles where the
dust lanes intersect the nuclear ring.
This can be compared with the
magnetic fields in the inner regions of our models (seen most clearly
in Figs.~\ref{model101}, \ref{modelunif}a, \ref{alpprom} where the
field in the inner regions is emphasized). The form of these fields is
remarkably similar to that shown in Fig.~\ref{N1097obs}.
We emphasize that the fact that
the observed polarized intensity does not vanish at small galactocentric
radii (see Fig.~\ref{radB2}), as it should for modes with even $m$, indicates
that the modes with odd $m$ should become increasingly dominant at smaller
radii. Angular momentum transfer due to the magnetic stress produced by the
regular field can provide a mass inflow rate into the central region of about
$1\,M_\odot\yr^{-1}$,  which is comparable to that required to feed the
active nucleus (Beck et al.\ 1999).

\section{Conclusions}
Our simulations confirm that conventional mean-field dynamo models
can reproduce magnetic field
structures that are similar to those observed in barred galaxies.
Even though we have used a velocity field model for a generic barred
galaxy (Athanassoula 1992), our magnetic configurations show an overall agreement
with observations of the prototypical barred galaxy NGC~1097.
Our models are most successful when they include enhanced turbulence in regions
of strong shear, most notably near the shock fronts offset from the bar major
axis.  The best overall agreement with observations seems to be provided by
models with $\widetilde\alpha\propto\omega$ and $f_\eta\approx3$, although
the overall appearance of the field in the outer regions is not very sensitive
to the choice of parameters.

We argue that regular
magnetic fields can be dynamically important in large regions
within the corotation radius. The energy density of the magnetic field
in our models and, plausibly, in real barred
galaxies exceeds that of the interstellar turbulence and is largely
controlled by the local shear in the regular (noncircular) velocity.
This is in a striking contrast to the situation in normal spiral
galaxies where regular magnetic fields and turbulence are close to
energy equipartition.

Our models can successfully reproduce
some salient features of the observed field structures, notably
the unexpectedly strong field upstream of the shock, the gentle
bending of the field across the shock, and the
overall structure of the field near the galactic centre.
There are well developed trailing spiral arms present
outside of the corotation radius, which begin approximately at the
ends of the bar.

The intensity of interstellar turbulence must be enhanced in the dust
lanes and circumnuclear region in order to obtain a radial contrast in the
regular magnetic field similar to that observed in NGC~1097. We argue
that this enhancement can arise naturally due to shear flow
instabilities.
The enhancement of turbulence may have important implications for star
formation
and gas dynamical models in barred galaxies.
Enhancements in turbulent velocity implied by CO observations for
the central parts of NGC~1097 (Gerin et al.\ 1988) and NGC~3504
(Kenney et al.\ 1993) are consistent with those used in our models.

The initially surprising bending of magnetic lines upstream of the
shock can be explained by (i) advection of magnetic field from other
regions,
 and (ii) by the alignment of $\vec B$ with the
principal axis of the rate of strain tensor rather than the velocity
itself.  Note however that advection in an approximately uniform
(non-sheared) flow does not cause alignment of $\vec{u}$ and
$\vec{B}$, even in the perfect conductivity limit.

In conclusion, we emphasize three important findings of this paper.
\begin{enumerate}
\item Barred galaxies are different magnetically to `normal' spiral
galaxies, in that the field strengths are significantly higher, and
the alignment between magnetic and velocity vectors is, on average,
closer.

\item The magnetic field may be dynamically important: Alfv\'en speeds can be
comparable with noncircular velocities (although generally a little lower).
Such strong fields require relatively, but not unrealistically,
strong dynamo action ($R_\alpha\ga 3$ for our models with $\eta_0 = 10^{26}$ cm$^2$s$^{-1}$).

\item In order to model satisfactorily the magnetic fields it is necessary for
the turbulence to be locally enhanced in the vicinity of the dust lanes and the
circumnuclear ring, in a manner consistent with other evidence. This will
affect other aspects of modelling of these galaxies.
\end{enumerate}

\begin{acknowledgements}
We are indebted to  E.~Athanassoula for providing us with the velocity
and density data from her model. We also thank
{\bf D.~Elstner,}  P.~Englmaier and
M.~Urbanik for useful discussions and comments.  We acknowledge
support from PPARC, NATO (grant PST.CLG 974737) and the RFBR (grant
01-02-16158). The generous assistance of W.~Dobler is gratefully acknowledged.
\end{acknowledgements}


\end{document}